\def\to{\hbox{$\,$--$\,$}}
\def\minus{\hbox{$\,$--$\,$}}
\def\muspc{\hskip 0.15 em}

\def\mag{\hbox{$\;.\!\!\!^m$}}
\def\etal{\mbox{et~al.}}
\def\AMORE{\textsf{AMORE}}
\def\HRDGST{\textsc{\tt HRD-GST}}
\def\HRDZVAR{\textsc{\tt HRD-ZVAR}}
\def\POWELL{\textsc{\tt POWELL}}
\def\PIKAIA{\textsc{\tt PIKAIA}}
\def\ZVAR{\textsc{\tt ZVAR}}
\documentclass{aa}
\usepackage{graphicx,color,longtable,amssymb}
\begin{document}

\title{Automatic Observation Rendering (\AMORE)}
\subtitle{I. On a synthetic stellar population's colour-magnitude diagram}

\author{Y.K. Ng\inst{{1,2}} 
\and{E. Brogt\inst{1,3}}
\and{C. Chiosi\inst{4}}
\and{G. Bertelli\inst{{2,5}}}}

\offprints{Yuen K. Ng}

\institute{TNO TPD Space, P.O. Box 155, 2600 AD \ Delft, the Netherlands
({\tt ykng\char64tpd.tno.nl})
\and{Padova Astronomical Observatory,
Vicolo dell'Osservatorio 5, I-35122 Padua, Italy
({\tt bertelli\char64pd.astro.it})}
\and{Kapteyn Astronomical Institute, P.O Box 800, 9700 AV Groningen,
the Netherlands (\tt brogt\char64astro.rug.nl)}
\and{Padova Department of Astronomy,
Vicolo dell'Osservatorio 2, I-35122 Padua, Italy 
({\tt chiosi\char64pd.astro.it})}
\and{National Council of Research, IAS, Rome Italy}
}

\date{Received February 14, 2002 / Revised April 10, 2002 / Accepted May 15, 2002}

\markboth{Y.K. Ng et~al.: \AMORE\ on a synthetic stellar population}
{Y.K. Ng et~al.: \AMORE\ on a synthetic stellar population}

\abstract{A new method, \AMORE\ - based on a genetic algorithm optimizer, 
is presented for the automated study of colour-magnitude diagrams.
The method combines several stellar population synthesis tools developed
in the last decade by or in collaboration with the Padova group. 
Our method is able to recover, within the uncertainties,
the parameters 
-- distance, extinction, age, metallicity, index of a power-law 
initial mass function and the index of an exponential star formation rate --
from a reference synthetic stellar population.
No {\it \`a priori}\/ information is inserted to recover 
the parameters, which is done simultaneously and not one at a time.
Examples are given to demonstrate and to better understand 
biases in the results, if one of the input parameters is deliberately 
set fixed to a non-optimum value.
\keywords{methods: data analysis, numerical --- Stars: HR-diagram, 
statistics}}

\maketitle
\noindent
\thanks{The appendix and the tables 
(electronically available at the CDS)
listed therein are 
published in the electronic version of this paper.}

\hyphenation{SDSS}
\hyphenation{RGB}
\section{Introduction}
\label{Introduction}
The present data flow of many ongoing surveys ---
such as 2MASS (Beichman et~al.\ \cite{2MASS98},
Skrutskie \cite{Skrutskie98}), 
DeNIS (Epchtein et~al.\ \cite{DeNIS97}, \cite{DeNIS99}), EIS 
(Renzini{\muspc\&\muspc}da~Costa~\cite{RenzinidaCosta97},
da~Costa~\cite{daCosta97}, da~Costa~\etal~\cite{daCosta98ea}),
OGLE-II (Udalski et~al.\ \cite{OGLEII}, Paczy\'nski 
et~al.\ \cite{Paczynski99ea}), 
SDSS (Fan \cite{Fan99} and references cited therein),
and even upcoming surveys as 
GAIA (Gilmore et~al.\ \cite{Gilmore98ea}, Perryman et~al.\ 
\cite{Perryman01ea}) ---
is so large that one requires either a semi-automated or a fully automated
method to analyse the colour-magnitude diagrams (CMDs) in the 
resulting databases. In this paper we discuss the 
development and the tests of an automated analysis method, which fully
employs the colour and magnitude information 
available about the stars populating the CMD.
Our method is based on an implementation of the 
CMD diagnostics suggested by Ng (\cite{Ng98a}). The method uses, in
contrast to other techniques (see Bertelli et~al.\
\cite{Bertelli92ea}, Gallart et~al.\ \cite{Gallart96ea} \&
\cite{Gallart99ea}, Geha et~al.\ \cite{Geha98ea}
{ Harris \& Zaritsky \cite{HarrisZaritsky2001},
Hernandez et al. \cite{Hernandez99ea}
}, 
Holtzman et~al.\ \cite{Holtzman97ea} \&
\cite{Holtzman99ea}) 
the full, { unbinned }
distribution of magnitudes and colours of the stars populating the CMD.
\par
The purpose of this paper is to verify 
that astrophysical parameters for a synthetic single stellar population
can be reliably retrieved with the so-called `{\sf A}\/uto{\sf M}\/atic
{\sf O}\/bservation {\sf RE}\/nderer' \AMORE.
In Sect.~\ref{AMORE} an outline of \AMORE\  is given 
together with its individual building blocks.
In Sect. ~\ref{Method} we outline the method we use and in
Sect.~\ref{test} we describe the tests performed with synthetic
stellar populations. The results are given in Sect.~\ref{Results} and
we discuss in Sect.~\ref{Discussion} the practical limits on the convergence,
which is imposed by some degeneracy of the parameter space.
We end with prospects on forthcoming tests, 
recommendations for improvements, and an outlook on future 
developments.

\section{\AMORE}
\label{AMORE}

\subsection{Project outline}
\label{ProjectOutline}
\AMORE\ tries to find the best matching synthetic CMD to an observed CMD. 
Such a synthetic CMD contains for stellar aggregates the contribution
of one or more stellar populations at the same distance. In the case
of a CMD along a particular line of sight in our Galaxy the synthetic
CMD can moreover contain the contribution of various populations with
stars distributed at different distances.
\hfill\break
\noindent
In this paper we focus on the implementation and the performance 
of \AMORE\ for the fitting of CMDs. For sake of argument only one,
single stellar population has been considered. 
The implementation of automatic fitting CMDs with multiple stellar
populations with stars at the same or different distances will
be subject of forthcoming papers.\\

\subsection{Building blocks}
\label{BuildingBlocks}
\begin{figure*}
\resizebox{10cm}{!}{\includegraphics{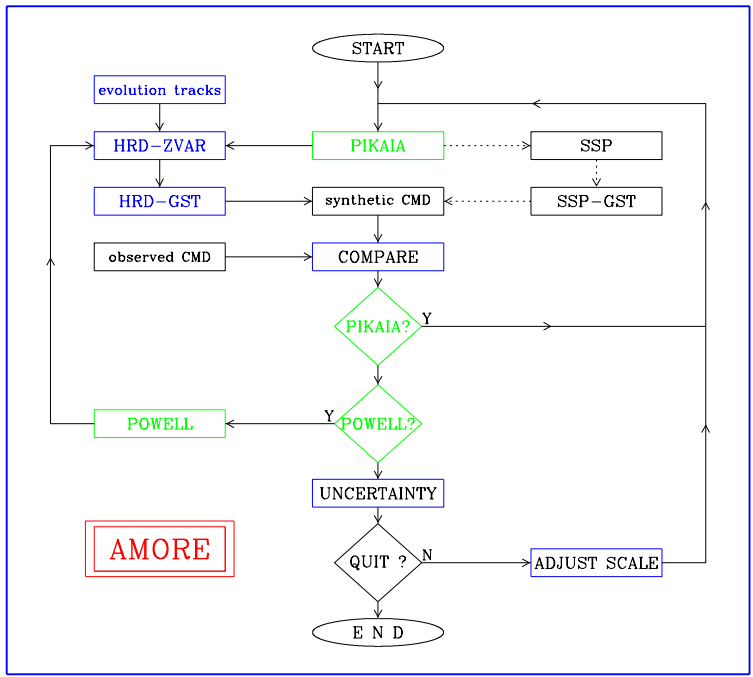}}
\hfill
\parbox[b]{75mm}{%
\caption{\AMORE\  flowchart: schematic diagram of the individual building 
blocks. \PIKAIA\ outlines the direction of the evolution paths
to be investigated.
Input for the stellar population synthesis engine \HRDZVAR\ is
the Padova library of evolutionary tracks.
The luminosity and effective temperature for each synthetic star of 
arbitrary metallicity is then transformed to an absolute magnitude 
in a photometric passband with the method outlined 
by Bertelli et~al.\ (\cite{Bert94ea}) and Bressan et~al.\ (\cite{Bres94ea}). 
The synthetic HRD is then `observed' and `detected'
through a Monte Carlo `observing run' with the \HRDGST.
Note that an alternative route is possible with 
single stellar populations (SSPs).
The synthetic CMD is compared with the observed diagram
and a fitness parameter is subsequently communicated to \PIKAIA,
which suggests a new set of parameters for each trial.
The iteration lasts for a user defined, fixed number of trials.
\POWELL's method of minimization is subsequently applied to get closer 
to the local or global minimum.
After computation of the uncertainties for each parameter the
evolutionary run is either aborted or a new \PIKAIA\ cycle is
started after shrinking the limits of the parameter space 
(see Sect.~\ref{contract}) 
}
\label{amorefloat}}
\end{figure*}

\AMORE, see Fig.~\ref{amorefloat} and Sect.~\ref{evolution} for details,
combines various analysis tools developed and improved at Padova
during the last decade. Conceptually, it is made up out of the
following building blocks: 
\begin{itemize}
\item{} a synthetic Hertzsprung-Russell Diagram generator
(hereafter referred to as \HRDZVAR\, see Sect.~\ref{ZVAR}),
\item{} a mockup version of the HRD Galactic Software Telescope (\HRDGST,
see Sect.~\ref{GST}),
\item{} a statistical diagnostic tool to compare the observed and synthetic 
CMDs with each other (Ng \cite{Ng98a}), and 
\item{} the \PIKAIA\ (extended version) and \POWELL\ optimizers (see
respectively Sects.~\ref{PIKAIA} and \ref{POWELL}), which search for
the best fit between an observed and a synthetic CMD in a
multi-parameter space.
\end{itemize}
In the following subsections a description is given of the various
building blocks.

\subsection{\HRDZVAR}
\label{ZVAR}
In the late-eighties Bertelli developed a code to generate synthetic
Hertzsprung-Russell diagrams (HRDs) from the isochrones computed by
the Padova group (cf. Chiosi et~al.\ \cite{Chiosi89ea}). Initially the
synthetic colour-magnitude diagram (CMD) technique was applied mainly
in the studies of LMC open clusters\footnote{The analysis technique
with stellar ratios was employed. The reason for this is that ratios
are less sensitive to uncertainties in certain regions in the CMD,
which might not be reproduced properly due to various reasons such as
the input physics used for the calculations of the stellar
evolutionary tracks or the transformations from the theoretical to the
observational plane.} (see for example Bertelli et~al.\
\cite{Bert85ea}, \cite{Bert90ea} or Chiosi et~al.\ \cite{Chiosi89ea}),
through which the amount of convective overshoot was calibrated
for the computation of a new generation of stellar evolutionary tracks.
Successive improvements were gradually applied when new sets of
evolutionary tracks (see Bertelli et~al.\ \cite{Bert94ea} for details)
were computed with improved radiative opacities (Iglesias et~al.\
\cite{Iglesias92ea}).
\par
The backbone of \HRDZVAR, the extended version of the HRD generator, is formed by the evolutionary tracks 
computed by Bertelli et~al.\ (\cite{Bert90ea}; \mbox{Z\muspc=\muspc0.001}),
Bressan et~al.\ (\cite{Bres93ea}; \mbox{Z\muspc=\muspc0.020}), 
and Fagotto et~al.\ (1994abc; 
\mbox{Z\muspc=\muspc0.0004}, 0.004, 0.008, 0.050, 0.10).
The metallicities of the tracks follow the enrichment law 
\mbox{$\Delta${Y}/$\Delta${Z}\muspc=\muspc2.5}
(see references cited in Chiosi 
\cite{Chiosi96} and Pagel{\muspc\&\muspc}Portinari \cite{PP98}).
\par
\HRDZVAR\ indicates that the metallicity Z is not limited to the
fixed values for which the evolutionary tracks have been computed, but
is variable through interpolation between the metal-poorest and metal-richest
tracks available inside the database of evolutionary tracks. In this
way one is able to generate synthetic stellar populations with a
smooth metallicity coverage. A prerequisite however is to use a
complete and homogeneous library of evolutionary tracks and some
improvements are expected if one adopts a grid of tracks with a
smoother metallicity coverage.
\par
\HRDZVAR\ has been distributed (privately) to various research groups
and is also referred to as \ZVAR\footnote{Note that an older and
modified version of \HRDZVAR\ is actually the program used to generate
the Bertelli et~al.\ (\cite{Bert94ea}) isochrones. \HRDZVAR\ is free of
the interpolation difficulties as reported by Olsen (\cite{Olsen99})}.
The version distributed, modified, and used by for example Aparicio
(\cite{Aparicio99}), Gallart (\cite{Gallart98}), and Ng et~al.\
(\cite{Ng96ea},\cite{Ng97ea}) is from now on referred to as
V1.0. Version V1.6 is used for the simulations and results presented
in this paper. This version contains a number of modifications and
improvements which speed up the code and fix some (rarely encountered)
bugs which interfered with the automatic minimization
process. Although various analysis methods are available, we limit
ourselves here to the description of the parameters related to the
\HRDZVAR\ as adopted for \AMORE.
\par
After selection of a set of tracks with fixed metallicity
and the choice of the parameters $\eta_{RGB}$ and $\eta_{AGB}$,
(the mass loss along the Red Giant Branch (RGB) and the Thermally
Pulsing Asymptotic Giant Branch (TP-AGB) phases respectively), the
major input to be specified for \HRDZVAR\ are:
\begin{itemize}
\item the metallicity range, Z ranges from
Z$_{min}$ to Z$_{max}$; 
\item the age range, the age ranges from 
$t_{min}$ to $t_{max}$; 
\item the slope $\alpha$\/ for the power-law IMF (Initial Mass Function); and
\item the index $\beta$\/ for the exponential{\footnote{{The SFR is only fixed 
for the time being to the adopted exponential shape}}}
SFR (Star Formation Rate). 
\end{itemize}
The luminosity and effective temperature for each synthetic star of 
arbitrary metallicity is transformed to an absolute magnitude 
in a photometric passband with the method outlined 
by Bertelli et~al.\ (\cite{Bert94ea}) and Bressan et~al.\ (\cite{Bres94ea}). 
Default setup for \HRDZVAR\ is the UBVRI JHKLMN\footnote{Note that the
IR photometric system is based on an `average' photometric system as
described by Bessell\mbox{\muspc\&\muspc}Brett (\cite{BB88}). Proper
transformations ought to be applied to the actual photometric system
prior to any astrophysical interpretation of the results.} broadband
photometric system. The setup can be altered to mimic any system,
given the description of the spectral response of the filter and the
detector of the photometric system.

\subsection{\HRDGST}
\label{GST}
\HRDZVAR\ was integrated in a galactic model by 
Ng (\cite{Ng94}, 1997ab). The
\HRDGST\ (Galactic Software Telescope) has been applied in the
studies of the galactic structure towards the Galactic Centre
(Ng et~al.\ \cite{Ng95ea}, \cite{Ng96ea} and 
Bertelli et~al.\ \cite{Bert95ea}, \cite{Bert96ea}) 
and other regions in our Galaxy (Ng et~al.\ \cite{Ng97ea}).
In this paper we do not require the full complexity of the structural
properties from the GST model. We only use a limited number of options
to `observe' a synthetic HRD at the suggested distance and to
simulate the photometric errors, extinction and crowding. 
\par
A table of the photometric errors, covering a specific magnitude interval 
per passband, is used and the program interpolates 
linearly to obtain the intermediate values. We assume for the 
simulation that the photometric errors are Gaussian distributed.
{A different description of the photometric
errors will be used when published artificial star tests 
(Stetson and Harris \cite{StetsonHarris88}, Gallart et al. \cite{Gallart99ea})
on an observational data set are indicative for 
a significant deviation from a Gaussian behaviour.}
\hfill\break
The visual extinction is simulated with the average value provided
and appropriately scaled to a value in different passbands. 
In the UBVRI passbands we adopted the scaling according 
to van~de~Hulst (\cite{vandeHulst49}; curve no.~15)
and for the JHKLMN passbands we follow the scaling laws
provided by Rieke\mbox{\muspc\&\muspc}Lebofsky (\cite{RiekeLebofsky85}). 
We do allow for some random scatter around the average extinction.
However, we do not consider (yet) the effects due to patchiness of the
extinction along the line of sight. Ng{\muspc\&\muspc}Bertelli
(\cite{NB96}) demonstrated that this is
in first approximation, 
visually almost indistinguishable from a random scatter around an
average extinction. 
\par
In many studies the observations are crowding limited, due to the
increasing number of stars towards fainter magnitudes. Crowding gives
rise to star blends which affects the magnitude and the colour of the
stars. The group of stars will be detected as a single star with a
magnitude equal to the sum of the stellar flux of the stars involved
in the stellar blend. The remaining stars are `hidden' from
detection. \\
The blends are well described as unresolved, apparent
binaries. The simulation of apparent binaries is made with an
iteratively improved blending probability, which is defined as the
probability that a star within a given ensemble of stars might blend
with another star from the same population. Each synthetic star within
a stellar population is tested against the blending probability. 
\\
The percentage of artificial binaries is with this scheme about twice the
blending factor. The blending factor in different passbands is not
necessarily the same and the occurrence of star blends is furthermore
not necessarily correlated, due to possible differences in the
exposure time or seeing conditions.
\\
{
The fainter companion stars of artificial binaries will give rise
to incompleteness of the synthetic stellar sample. This allows us
to map the synthetic stellar completeness function, which
can be compared directly with the completeness function obtained from 
artificial star tests from the observed stellar sample. }
\par
In the CMD to be analysed we assume implicitly that,
between the observed and synthetic photometric system, 
the uncertainty in the magnitude zeropoint is smaller than 
the uncertainty in the zeropoint of the colour,
see Carraro et~al.\ (\cite{Carraro99ea}), and references cited therein.
We allow for this reason the possibility that a small zeropoint
shift might be present between the colours of these systems.
\begin{figure*}
\resizebox{8.7cm}{!}{\includegraphics{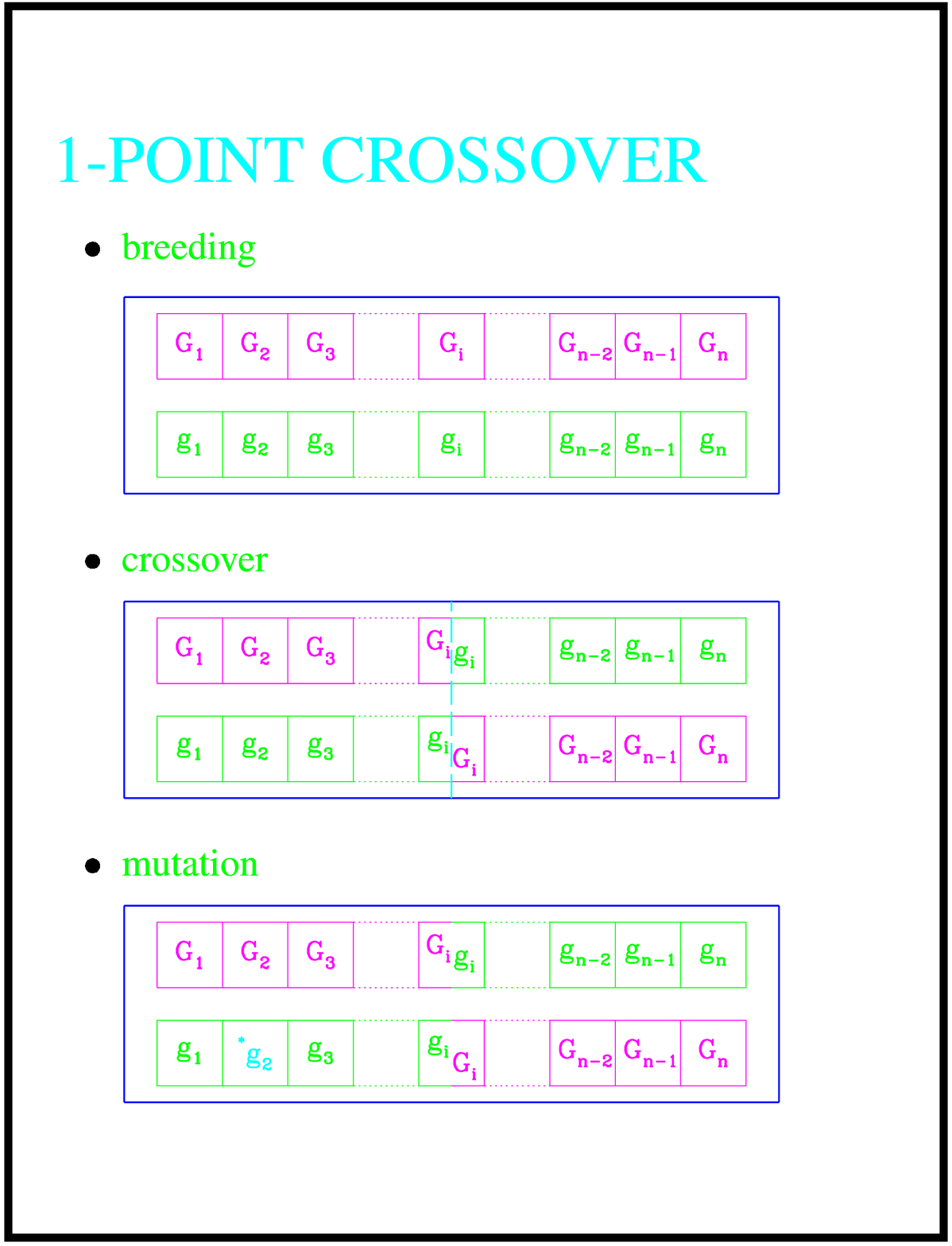}}
\hfil
\resizebox{8.7cm}{!}{\includegraphics{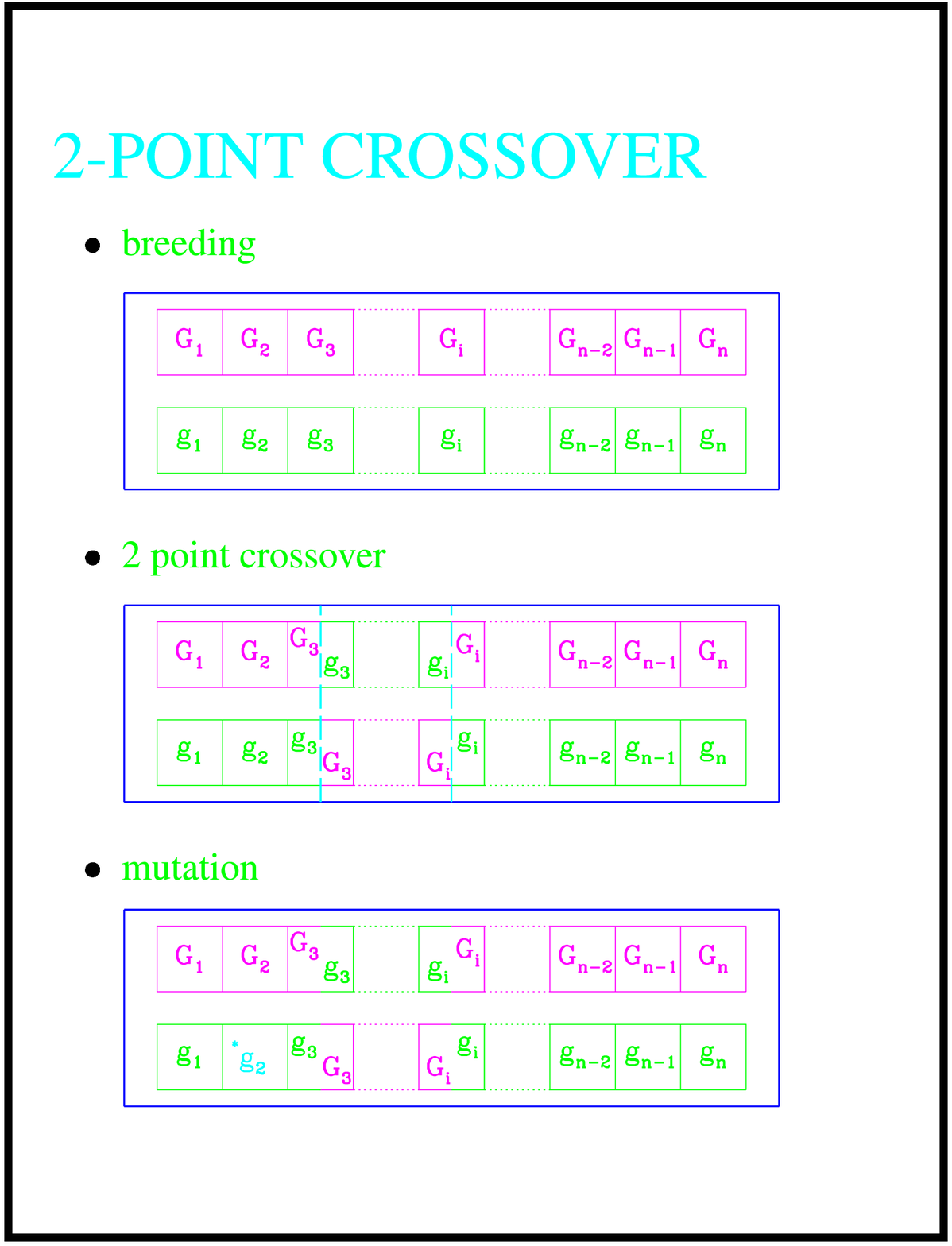}}
\hfil
\vbox{
\caption{Schematic view of the main genetic operators 
acting on two sets of parameter strings 
in order to procreate two new sets {(one chromosome set for each individual) }
of test strings: 
natural selection of two individuals
{\color{magenta}$G$}\/ and {\color{green}$g$} for the breeding of offspring, 
modification of the genetic content by means of a 
homologous crossover -- 
at gene ${\color{magenta}G_i}/{\color{green}g_i}$, 
and finally the mutation of gene {\color{green}$g_2$} to 
{\color{cyan}$^*g_2$}. 
{Each gene on a chromosome represent a parameter to be 
optimized. }
The schematics for a 2-point crossover scheme are shown for comparison
\label{crossoverfig}}}
\end{figure*}

\subsection{SSPs}
\label{SSPs}
Although straightforward, the generation of a large number of
synthetic HRDs with each their own specific age-metallicity range,
star formation history and initial mass function can be a time
consuming task, because of the repetition of many calculations to
generate one diagram. Figure~\ref{amorefloat} indicates that an
alternative route with single stellar populations (SSPs) is available
for the automated analysis. However, this method requires the
computation of a large, regular grid of SSPs for different
age-metallicity ranges and IMFs. The star formation history and
age-metallicity range are the result of the linear combination of
the SSPs. The time spent on computations
of new CMDs thus can be greatly reduced
through the use of probability density diagrams. 
However, this is not our prime objective.
Our present goal is to develop an automated fitting method,
which compares on a star by star basis, 
and to demonstrate its potential. Optimization for speed 
is not yet our primary concern.
Moreover, the generation of one synthetic CMD with 5000 stars
takes about 0.75s on a PC equipped with a 200~MHz Pentium processor.
This is a good indication that our present version of the software
tool is performing at an acceptable speed. We refer to Dolphin
(\cite{Dolphin97, Dolphin01etal, Dolphin02}) 
and Olsen (\cite{Olsen99}) for a description of a
method using SSPs.
\par
\begin{table*}[t]
\caption{\PIKAIA\ control parameters. These parameters were kept fixed
at the values listed throughout the evolutionary runs
\label{pikaiasetup}}
\begin{center}
\begin{tabular}{|lll|l|}
\hline
initial & \PIKAIA & parameter & comment\\
 value & default     & \ identifier & \\
\hline
\hline
100	& 100   & np	 & the number of individuals per population\\
20	& 500	& ngen	 & number of generations\\
2	& 5	& nd	 & number of digits encoding accuracy\\
2	& 2	& imut	 & mutation mode, imode=2 then pmut=[pmutmn,pmutmx]\\
0.005	& 0.005	& pmut	 & initial mutation rate\\
0.005	& 0.005	& pmutmn & minimum mutation rate\\
0.35	& 0.25	& pmutmx & maximum mutation rate\\
0.95	& 1.0	& fdif	 & fitness differential\\
3	& 1	& irep	 & reproduction plan\\
0	& 1	& ielite & elitism\\
0	& 0	& ivrb	 & verbose mode\\
\hline
\end{tabular}
\end{center}
\end{table*}

\subsection{\PIKAIA}
\label{PIKAIA}
The \PIKAIA\ optimization package (V1.0; public domain) was developed by
Charbonneau (\cite{Char95}) and a full description of this package is
given by Charbonneau{\muspc\&\muspc}Knapp (\cite{CK96}). 
\PIKAIA\ has been used successfully in a wide range of astrophysical
applications (e.g. Bobinger \cite{Bobinger2000}, Charbonneau et~al.\
\cite{Charbonneau98ea}, Gibson \& Charbonneau
\cite{GibsonCharbonneau98}, Kaastra et~al.\ \cite{Kaastra96ea},
Kennelly et~al.\ \cite{Kennelly96ea}, Lamontagne et~al.\
\cite{Lamontagne96ea}, McIntosh et~al.\ \cite{McIntosh98ea}, Metcalfe
\cite{Metcalfe99}, Mewe et~al.\ \cite{Mewe96ea}, Noyes et~al.\
\cite{Noyes97ea}, Saha \cite{Saha98}, Wahde \cite{Wahde98}).\hfill\break
\PIKAIA\ is based on a genetic algorithm (Holland \cite{Holland75},
Goldberg \cite{Goldberg89}, Davis \cite{Davis91}, De Jong
\cite{deJong93}), and is in principle not a function optimizer, but it
does this extremely well. It searches for, locks on to, and pins down
an optimal solution in a way, which is conceptually comparable to
biological evolution through natural selection. Genetic algorithms are
capable to explore and find in a robust way an optimum, but not
necessarily the best, setting for a particular problem. 
{ In our case this comes down to minimizing the difference
between a synthetic and an observed CMD by evolving the astrophysical
parameters that define the shape of the CMD (see Table~\ref{amorehrd}
and Sects~\ref{fparameters} and \ref{Method}). }
We follow the
generally accepted  biological terminology for the description of a
genetic algorithm.
\par
A genetic algorithm makes use of a reduced version of 
the evolutionary process. The gene pool, 
{ i.e. the set of parameters to be optimized, }
and its associated phenotypic population evolves in response to 
\begin{itemize}
\item the reproductive success of the population, following its
reproduction plan;
\item genetic recombination (crossover, 
{see Fig.~\ref{crossoverfig}}) at breeding;
\item random mutations 
{(see Fig.~\ref{crossoverfig}) }
during breeding, which affect a subset of the events.
\end{itemize}

\subsubsection{Flow control}
\label{flow}
\PIKAIA\ has 12 flow control parameters which are discussed 
by Charbonneau (\cite{Char95}) to whom we refer
for a detailed description. We limit ourselves
to a short summary and the purposes of these control parameters:
{\tt np} defines the number of individuals in a population;
{\tt ngen} specifies the number of generations that the population 
is evolving;
{\tt nd} is number of digits encoding 
accuracy{\footnote{{The 1 or 2 number of digits encoding accuracy 
maps the parameter via an integer value, either 0\to9 or 0\to99, 
to a range of floating point values by a controlling `normalization'
function, see Charbonneau (\cite{Char95}) for additional details
}} }
used for the parameters; 
{\tt pcross} is the probability that a crossover occurs between
the chromosomes of the parents;
{\tt imut},
{\tt pmut}, {\tt pmutmn},
and {\tt pmutmx} specify the mutation mode, the mutation rate
(the initial mutation rate if the rate spans the range from
{\tt pmutmn} to {\tt pmutmx} for
{\tt imut\muspc=\muspc2}),
and the minimum and maximum mutation rate; 
{\tt fdif} the fitness differential controls the selection of 
the individuals for breeding through their fitness;
{\tt irep} defines the reproduction plan to be followed;
{\tt ielite} defines if the fittest individual can or cannot 
be selected for replacement; and
{\tt ivrb} specifies verbose mode for extra on screen 
information during the evolutionary run.
\hfill\break
Table~\ref{pikaiasetup} holds a list of 
the \PIKAIA\ flow parameters which were kept constant 
during all the simulations described in this paper. 
In the following subsections we describe the extensions added to
the 12~flow parameters of \PIKAIA.

\subsubsection{Crossover}
\label{crossover}
The crossover operator is very effective in a global exploration of
the full parameter space and is in a way comparable to a variational
calculus method. A one-point crossover scheme, see
Fig.~\ref{crossoverfig}, is sometimes inadequate to combine and
pass on certain features encoded on the chromosomes (Michalewicz
\cite{Michalewicz96}) to its offspring. In some cases only a
correlated modification of a number of genes, say 2, will result in a
fitter offspring. This can, for example, be mimicked through the
application of a two- or multi-point crossover scheme. \PIKAIA\ has
been extended with the control parameter {\tt rcross} to handle a
multi-point crossover operation. For example: {\tt
rcross\muspc=\muspc1} represents the default one-point crossover,
while {\tt rcross\muspc=\muspc2.3} represents a `2.3-point' crossover:
i.e. a two-point and three-point crossover for respectively 70\% and
30\% of the cases (2.3\muspc=\muspc0.7\muspc$\times$\muspc2\muspc+
\muspc0.3\muspc$\times$\muspc3).

\subsubsection{Brood recombination}
\label{brood}
An important drawback of genetic algorithms is that
the crossover operator is for about 75\% of the time 
lethal to its offspring, i.e. it produces children which
are not as fit as their parents (Banzhaf et~al.\ \cite{Banzhaf98ea}).
To avoid missing, potentially, fitter offspring and to
reduce the destructive effect of the crossover operator
we have incorporated brood recombination in \PIKAIA\ 
through a new control parameter {\tt rbrood}.
The default \PIKAIA\ reproduction scheme is obtained with 
{\tt rbrood\muspc=\muspc1}, i.e. two parents breed once and 
produce two new individuals.
With {\tt rbrood\muspc=\muspc3.5} the parents are allowed 
to breed on average 3.5 times to produce a larger offspring (in this
case 7 on average).
{For \mbox{\tt rbrood\muspc>\muspc1} one ends up with more
than two offspring. \PIKAIA\ on the other hand expects from each pair 
of parents only two children. This constraint was obeyed
in order not to significantly alter the global behaviour 
of \PIKAIA. Therefore, in order to avoid an exponential growth of 
the population for {\tt rbrood\muspc$\ne$\muspc1},
only the fittest two individuals of the local offspring 
survive\footnote{{ This is irrespective if the local, fittest 
individuals are weaker than the weakest individual in the global 
population or if the locally remaining offspring are fitter
than the fittest individual in the global population}}
and enter the global population for a fitness 
evaluation from which a selection is made for further breeding.
}
\\
The extra breeding increases the computational effort considerably,
due to a larger number of function evaluations. The advantage is a more
rapid increase of fitter individuals through the selection
of effective crossovers from good recombinations. On the other hand, a
rapid increase of fitter individuals might lead to a premature
convergence to a local minimum, due to a smaller variance in the genetic pool.

\subsubsection{Creep mutations}
\label{creep}
We further introduced the `creep' mutation  
(Charbonneau{\muspc\&\muspc}Knapp \cite{CK96}) in order to overcome
the so-called `Hamming Wall' problem, i.e.
the inability to cross in a decimal encoding scheme
certain boundaries with a one-point mutation operator.
The creep parameter {\tt pcreep} defines the probability that a gene in the pool
undergoes a `standard' mutation (change digit randomly in the range 0\to9)
or the `creep' mutation (add or subtract one from the current value of the
digit).
We adopted as default equal weight for the occurrence of a `creep' 
or `standard' mutation. 

\subsubsection{Correlated mutations}
\label{correlated}
In general mutations occur in the optimization process to avoid
premature convergence. A low mutation rate is sufficient for this
purpose. However, a high mutation rate can be used as an additional
way to explore the parameter space like a virus. Although one would
prefer to use a (multi-point) crossover operator, we do allow that
mutations can be used instead, simply because the two operators also
co-exist in nature.
\par
We modified \PIKAIA 's uniform mutation mode.
In the majority of the cases we require a \mbox{(anti-)}correlated change
between two or more parameters (see Sect.~\ref{convergence}).
In a standard mutation scheme convergence might be slow if one has to 
wait for the simultaneous occurrence of a favourable \mbox{(anti-)}correlated 
mutation of two specific parameters in order to improve the fitness.
We introduced an extra parameter {\tt pcorr} which defines the
probability that a correlated mutation occurs. 
If this is not the case the standard mutation scheme is chosen. 
Otherwise we allowed that in 50\% of the cases the mutations of two genes
({\tt igen1} and {\tt igen2} { are extra input parameters added to the modified 
version of \PIKAIA}) are more relevant 
than the mutations occurring in other parameters. 
For the remaining 50\% of the cases the two genes are determined
stochastically. Additional details about the adopted values of the control
parameters are given in Sect.~\ref{fparameters} and Table~\ref{first}. 

\subsection{Fitness}
\label{fitness}
\PIKAIA\ searches for the optimum solution
by maximizing the fitness function~$f$. 
The fittest solution has a fitness $f\!=\!1$, while the worst has 
$f\!=\!0$. We use the Ng (\cite{Ng98a}) fitness function, a
combination of a chi-squared and Poisson like functions. These
functions minimize the differences between observed and synthetic 
diagrams { via a star-by-star matching\footnote{{ This option is feasible due to the 
increment of the present day computational speed}} 
 scheme. 
}
\\
The Ng fitness function is defined as:
\begin{equation}
\quad f = {1\over{1+F}} \;, \label{eq fitness}
\end{equation}
where $F$\/ is
\begin{equation}
\quad F = F_\chi^2 + F_P^2 \;.\label{eq merit}
\end{equation}
$F_\chi$ is the chi-squared function of the
best fitted points within a \mbox{3\to5\muspc$\sigma$} error ellipse
and $F_P$ is the Poisson function of the residual points outside this 
ellipse. Both $F_\chi$ and $F_P$ are dimensionless, but they hold
information about the average uncertainty in units of $\sigma$,
say $\sigma_\chi$ and $\sigma_P$. For example: the average uncertainty
per point for a fit with $F_\chi$ is $F_\chi\times\sigma_\chi$.
\\
$F_\chi$ and $F_P$ are respectively defined as:
\begin{equation}
\quad F_\chi = {\sqrt{\overline{\chi^2}}} = \sqrt{\chi^2(O,S)/N_{match}} \;,
\end{equation}
and
\begin{equation}
\quad F_P = {{N_{O,not} + N_{S,not}}\over{\sqrt{N_O}+\sqrt{N_S}}}\;.
\end{equation}
The intuitive motive behind this is to make a division between 
the synthetic points matching the observed 
CMD for which the errors are expected to be normally 
distributed and the points which do not match
and are allegedly assigned to the Poisson merit function.
The method actually uses $F_\chi$ as a loosely fixed `anchor',
puts the outlier points in $F_P$ and then
reduces the number of unmatched points by
minimizing $F_P$.
We refer to Ng (\cite{Ng98a}) for additions details and a discussion
of these functions. Suffice to say that for an acceptable solution both 
$F_\chi$ and $F_P$ are about 1 or 
smaller\footnote{For our testcase, as described in Sect.~\ref{test},
$F_\chi$ ranges from 0.7\to1.0 for all observed data points with a matching 
synthetic point within a 3$\sigma$\/ uncertainty ellipse.
This corresponds to a goodness
of fit parameter ranging from 0.49\to1.0. It
further indicates that it is justified to assume 
that the measurement errors are normally distributed.}
which on its turn can be relaxed to the condition 
$F\!\la\!2\;$\footnote{This condition is comparable
but not equivalent to the results obtained by
Gallart et~al.\ (\cite{Gallart99ea}). They
demonstrated from a comparison with colour-magnitude bins
that a good agreement between the input and recovered
SFR(t) required a 
reduced chi-squared of $\chi_\nu^2\!\simeq\!2.0$.} 
and thus yields ${1\over3}\!\la\!f\!<\!1$.
\begin{table*}[t]
\caption{\AMORE\ control parameters, used as the `educated 
guess' for the astrophysical parameters 
in the first \PIKAIA\ cycle. These values result in a fitness
$f=0.004$. The value for the parameters are constrained 
between the lower and upper limit
\label{amorehrd}}
\begin{center}
\begin{tabular}{|lll|ll||l|}
\hline
\#&parameter & value & lower limit & upper limit & description\\
\hline
\hline
1 & log d& 3.7    & 3.6       & 4.0     & log of distance\\
2 & A$_V$ & 0.30   & \minus0.001    & 0.5     & extinction\\
3 & $\log t_{low}$& 9.5    & 9.0       & 10.3    & log lower age limit\\
4 & [Z]$_{low}$& \minus0.90  & \minus1.69897  & 0.69897 & log lower metallicity limit\\
5 & $\alpha$ & 1.35   & 1.001     & 3.5     & IMF slope\\
6 & $\log t_{high}$& 9.6    & 9.0       & 10.3    & log upper age limit\\
7 & [Z]$_{hgh}$ & 0.30   & \minus1.69897  & 0.69897 & log upper metallicity limit\\
8 & $\beta$& 1.0    & \minus2.00     & 5.00    & SFR slope\\
\hline
\end{tabular}
\end{center}
\end{table*}
\par
The formal 1$\sigma_k$ uncertainty of each parameter $k$,
see Table~\ref{tbl1},
is obtained through variation of this parameter and 
by minimizing the function $|\sqrt{F_k}-\sqrt{F_{min}}-1|$. 
Conceptually, this is similar to moving the merit function
in the $F_P,F_\chi${\to}plane away from its optimum setting,
to the nearest position on a contour +1$\sigma_k$ higher. 
The associated fitness
function $f_{\sigma,k}$\footnote{For both age and metallicity ranges the 
associated uncertainties denote for the lower and upper values
the $-1\sigma$ and $+1\sigma$ boundary} is: 
\begin{equation}
\quad f_{\sigma,k} = {1\over{1+|\sqrt{F_k}-\sqrt{F_{min}}-1|}} \;,
\end{equation}
where $k$\/ is the particular parameter for which the 
uncertainty is estimated and $F_{min}$ is the global value 
obtained for the fittest population.

\subsection{\POWELL}
\label{POWELL}
We implemented a hybrid optimizer in which we use 
\PIKAIA\ to explore the parameter space and then use
\POWELL's minimization algorithm (Powell \cite{Powell64};
Press et~al.\ \cite{Numrecip86}) to pin down the nearest
local or global minimum through a direction set method which produces
$N$\/ mutually conjugate (non-interfering) directions.
For details and an excellent description
of this algorithm we refer to Press et~al.\ and references cited therein.
\\
{ A hybrid minimization strategy is used, because 
\PIKAIA\ is by definition not a function optimizer, but it 
tends to get close near a (local) optimum. \POWELL\ 
is used to get even closer to the (local) optimum. 
If we had landed in a local optimum then we needed \PIKAIA\
to jump out of it. The origin of our need for a hybrid
search strategy is comparable to the minimization problems
encountered by Harris \& Zaritsky (\cite{HarrisZaritsky2001}).
}

\subsection{Contracting parameter space}
\label{contract}
A full exploration of parameter space at one digit accuracy would take
considerable time (in the order of weeks), even though certain
forbidden combinations of parameters can be excluded {\it \`a priori}.
A full exploration at two or more digits accuracy
is nearly impossible due to present day computational limits. 
\par
We implemented a dynamic, scalable parameter range in our search for
an optimum set of parameters. The parameter range shrinks 
after each optimization cycle with \POWELL.
This leads to an improved accuracy in the results
with a fixed number of digits encoding accuracy. \\
{
The automated re-scaling of the parameter range improves the 
resolution of the exploration of the search grid. In addition, due to 
the re-scaling one may circumvent partial 
degeneracy of the parameters. 
}
\par
The global function $F(F_\chi,F_P)$ gives the global distance to the minimum
in terms of $\sigma^2(\sigma_\chi,\sigma_P)$.
If we have $n$ parameters then each parameter $k$\/ is in a simple approach
on average about $\sqrt{F/{n}}\,\sigma_k$ away from its optimum
value, 
{because we assume that
\begin{equation}
  F = \sum_{k=1}^n 
  ({{y_{k,sim}-y_{k,true}}\over\sigma_k})^2 \;, \label{eq average sigma}
\end{equation}
where $y_{k,sim}$ is the simulated value of parameter $k$\/ and
$y_{k,true}$ is the true value of parameter $k$.
The average offset per parameter $k$\/ is therefore 
\mbox{$\sqrt{F/{n}}$}.
}
The 
{new }
limits can then be set to \mbox{$\pm\sqrt{F/{n}}\,\sigma_k$}.
To balance the cancellation of errors, due to a negative correlation
between some parameters, we adopt a rather conservative approach by 
constraining the limits of each parameter $k$\/ to
$\pm3\times1.3{\sqrt{F/{n}}}\,\sigma_k$. In addition we add the condition
that 3\muspc$\times1.3{\sqrt{F/{n}}}\!>\!0.005$.

\subsection{Adjustable parameters: free \& fixed}
\label{fparameters}
\AMORE\ uses in essence two different sets of parameters as input.
One set contains program flow control parameters and
information about the observational data to be simulated:
the photometric errors \& crowding factors per passband,
a shift of the zeropoint of the colours due to a difference 
between the observed and synthetic photometric system,  
and the spread around the average extinction
value to account partly for the differential reddening in a field.
\par
\hyphenation{Sect}
\begin{figure*}[t]
\begin{center}
\resizebox{8.0cm}{!}{\includegraphics{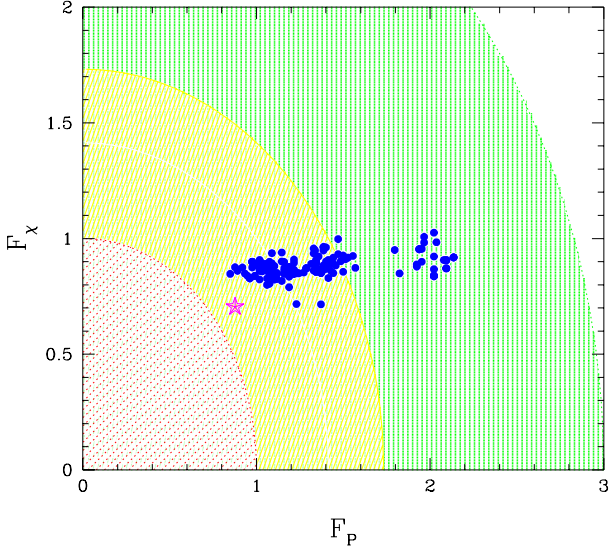}}
\quad
\vbox{\hsize=8.0cm%
\caption{Evolutionary status for all the trial models
(filled dots; see Table~\ref{first}
for details) for each population after 400 generations, see also
Fig.~\ref{evolveamorefig}.
The filled star {\color{magenta}$\bigstar$} points to the location
obtained with the original parameter settings as
given in Table~\ref{tbl1}.
Acceptable solutions are found in the region
for which both $F_\chi$ and $F_P$ are less than 1,
see Sect.~\ref{fitness}. The condition has been 
relaxed to $F\!<\!2$ (solid white line).\/
The asterisk indicates the position obtained for a different
realisation of the test population by changing the random seed. 
The shaded regions indicate solutions for which the 
difference between the CMDs from the 
`observed' and synthetic population is on average
less than respectively 3$\sigma$, $\sqrt{3}\sigma$, and 1$\sigma$.
Note that rounding errors, see Sects.~\ref{res_test1} \&
\ref{degeneracy},  
give rise to a degeneration of the parameter space 
around $f\!>\!0.25$ 
(i.e. 
$F\!<\!
{3}$
)
\label{generationfinfig}}\null\vfill\null}
\end{center}
\end{figure*}
The other set of input parameters is used by \PIKAIA\ and can be
divided into two parts (see Table~\ref{pikaiasetup} and
Table~\ref{amorehrd}). One part contains the \PIKAIA\ control
parameters. The other part contains the lower and upper limits of the
astrophysical parameters to be optimized as well as an initial guess
for the value of those parameters (first column of
Table~\ref{amorehrd}, resulting HRD in Fig.~\ref{hrdevolve}{\bf b}).
These parameters are a combination of the synthetic population's 
intrinsic properties, i.e. age, metallicity, slope of the power-law
IMF and the index of the exponential SFR. In our case we allow 
for age and metallicity not to be restricted to one fixed value,
but to cover a specific range (see Table~\ref{amorehrd}).
\hfill\break
In addition, there are two parameters which mimic the synthetic
population's behaviour as placed in a mockup version of our Galaxy.
These parameters are the distance from the Sun and
the average extinction.
\par
In total we thus have 8 free parameters which \AMORE\  has to optimize
simultaneously.

\section{Method}
\label{Method}

\subsection{Genetic evolution}
\label{evolution}
{
In terms of genetic programming the objective of \AMORE\ is to determine
the genome (i.e. the set of {\it astrophysical parameters}\/ 
described in Sect~\ref{fparameters}, see also Table~\ref{amorehrd}) 
of a specified individual (i.e. the {\it observed stellar population}). 
Note that it is not possible to directly observe the genome. 
The genome is determined from the phenotype of each individual
(i.e. the {\it synthetic CMDs}, see for example 
Fig.~\ref{hrdevolve}). The genetic information is located
in the genes of one chromosome\footnote{{
The genetic information is currently located on one chromosome.
Individuals with two chromosomes 
might be considered as a future extension. A two-chromosome approach
has the advantage that certain genetic information can remain present 
in a recessive form.} }
\par
\begin{figure*}[t]
\resizebox{18.0cm}{!}{\includegraphics{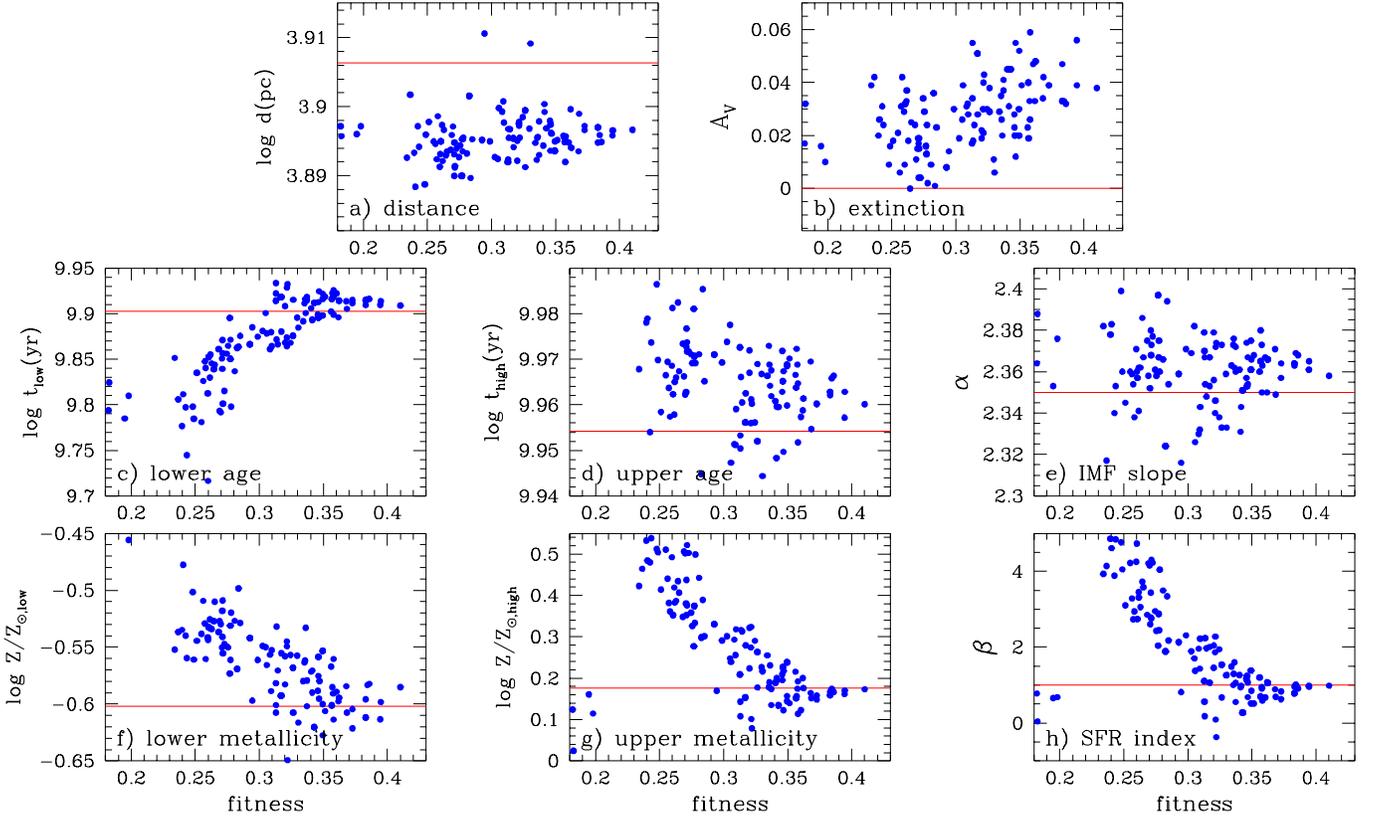}}
\caption{\label{evolveamorefig}
The filled circles in panels {\bf a}\to{\bf f} display the 
values obtained for the parameters using the models in Table~\ref{first}.
The solid line refers to the value set for the original population}
\end{figure*}
A guess of the genotype of the observed CMD is obtained 
through comparison with a synthetic CMD, which is generated 
via supervised evolution and breeding
(\PIKAIA\ together with \POWELL).
The stars in the synthetic CMD population with a particular genotype 
are raised to maturity (\HRDZVAR\ and \HRDGST).
A group of individuals\footnote{{ 
To avoid confusion the term {\it group}\/ is used instead of the
biological term {\it population}, 
because each individual in the group is actually a synthetic 
stellar population} }
is allowed to procreate (the chance of an
individual procreating depends on its fitness and the selection
pressure, see Charbonneau \cite{Char95} and Charbonneau \& Knapp
\cite{CK96}) and the genetic information of the parents is passed on
to their offspring (see Fig.~\ref{crossoverfig}).
\par
A fitness evaluation 
(a comparison between the observed and synthetic CMD)
provides a ranking of the resulting group of individuals. 
If the individual has 'good genes'
it survives, remains in the group 
and gets a chance of procreation.}\par
The evolutionary process of breeding and fitness
evaluation is repeated for a fixed number of generations.
The gene pool of the resulting best individual at the end of the
evolutionary run with AMORE hopefully represents a near-optimum
representation of the unknown genome.

\subsection{Running AMORE}
\label{running}
Initially \PIKAIA\ is in control (see Fig.~\ref{amorefloat}) of the
evolution for a fixed number of generations.
Afterwards \POWELL\ tries to improve the genome of the fittest individual
communicated through \PIKAIA. 
We then determine the uncertainty for each gene on the chromosome.
Subsequently, we tighten the limits on the range of variation allowed 
for each gene and re-scale the parameters on the genetic
print of the fittest individual accordingly.
During the shrinkage of the parameter range
we do not re-scale the genetic information of the remaining 
individuals, but preserve their former values as 
semi-random input for the continued optimization process.
{ The latter addition to the hybrid scheme
is most likely a significant driver in speeding up 
the search for a fitter individual.}
\hfill\break
After each optimization with \POWELL\ a new cycle with 
\PIKAIA\ is started with the current best parameter set as 
`educated next guess' for \AMORE's progressive evolution. 
The total number of \PIKAIA\ cycles is user defined.
\hfill\break
In Sect.~\ref{contract} we argued that
the parameters are on average about $\sqrt{F/n}\,\sigma_k$ away 
from its optimum value. 
The convergence however is not governed by the average `distance'
that each parameters is away from its optimum setting. It is mainly
determined from the ability to tune the parameter which has
the largest offset from its optimum value.
\hfill\break
In the \AMORE\  training sessions it was noted that with 
\mbox{$F\simeq\!3.0$} about three of the eight
parameters are about \mbox{1\muspc$\sigma_k$} 
(\mbox{$\simeq\sqrt{F/3}\,\sigma_k$}) 
away from their optimum value.  
\AMORE\  
{ has a built-in option to do }
a random variation from 0\to3~$\sigma_k$ of 
two parameters 
{(randomly selected) }
from the
running best solution when the fitness is less than 0.30.
Above this threshold we choose a new value 
for two parameters according $\pm\sqrt{F/\rho}\,\sigma_k$, where 
$\rho$ can be any number between 1.0 and 4.0.

\section{Tests}
\label{test}
\subsection{Test objectives}
We performed several tests on \AMORE\ in order to
\begin{itemize}
\item verify and validate \AMORE's performance in retrieving the
astrophysical parameters of a synthetic single stellar population; 
\item determine adequate values for the parameters
{\tt pcross, rcross, rbrood, pcreep, {\rm and} pcorr}
in the extended version of \PIKAIA;
\item study the effects of rounding and degeneracy;
\item study the effects of fixing parameters on the
convergence;
\item study the effects of a high extinction on the convergence.
\end{itemize}

\subsection{Setup}
\label{setup}
The hybrid interaction between \PIKAIA\ and \POWELL, combined with a
progressive shrinking/re-scaling of
the parameter space, requires that a trade-off has to be made in the 
choice of the size of the population and the number of generations 
we allow this population to evolve in order to obtain results in a
reasonable amount of CPU processing time.\\
We explored several different settings for the \PIKAIA\ control
parameters, because the tuning of those parameters is very problem
dependent (Charbonneau \& Knapp, \cite{CK96}). The values we decided
to use are listed in Table~\ref{pikaiasetup}. Four notes can be made
here.\\
Firstly
the steady-state-delete-worst reproduction plan ({\tt
irep=3}) we adopted, in which we replace the
least-fit individual from the population when the fitness of the 
new individual is superior to that of the least-fit population member.
Choosing this reproduction plan implies that the elitism
control parameter ({\tt ielite}) is non-operative, 
because elitism is active
by default. We evaluated two other
reproduction plans (Charbonneau \& Knapp \cite{CK96}); full
generational replacement and steady-state-delete-random. The
steady-state-delete-worst reproduction plan produced on average the
best results.\\
Secondly the mutation rate of 0.35
corresponds, in case of a default 2 digit accuracy, with the on
average occurrence of 2.8 mutations per astrophysical parameter.\\
Thirdly, the fitness differential parameter {\tt fdif}, 
a measure for the selection pressure, would normally be chosen as high
as possible ({\tt fdif=1} in this case). However, it
may possible to circumvent local minima by lowering that value a bit
(Charbonneau \& Knapp, \cite{CK96}). Setting {\tt fdif}=0.95 turned out
to be a good trade-off choice.\\
\begin{table*}
\vbox{\mbox{\vbox{\hsize=8.5cm%
\caption{Average fitness ($\overline{f_A}$) 
for different values of the parameters as
obtained from Table~\ref{first}.
The first column displays the parameter name which is
varied, the parameter value is given in the second column, 
the third \& fourth column show respectively $\overline{f_A}$,
uncorrected for {\tt pcorr=0.0},
together with its standard deviation, and
the fifth and sixth column display the averaged values
after removal of the results of  
the models with {\tt pcorr}=0.0.
See Sect.~\ref{detervalues} for additional details}
\label{avgfit}}}
\hfill
\mbox{\vbox{\hsize=8.5cm%
\begin{tabular}{|lc|cc|cc|}
\hline
parameter & value & $\overline{f_A}$ & $\sigma_{n-1}$ & $\overline{f_A}$ & $\sigma_{n-1}$\\
\hline
\hline
{\tt pcross} & 0.50 & 0.276 & 0.070 & 0.287 & 0.066\\
             & 0.85 & 0.299 & 0.063 & 0.297 & 0.056\\
{\tt rcross} & 1.00 & 0.293 & 0.065 & 0.290 & 0.063\\
             & 2.00 & 0.285 & 0.068 & 0.297 & 0.065\\
	     & 3.00 & 0.285 & 0.070 & 0.290 & 0.057\\
{\tt rbrood} & 1.00 & 0.286 & 0.077 & 0.230 & 0.059\\
	     & 2.00 & 0.292 & 0.058 & 0.301 & 0.055\\
	     & 4.00 & 0.285 & 0.066 & 0.276 & 0.067\\
{\tt pcreep} & 0.0  & 0.287 & 0.071 & 0.292 & 0.068\\
	     & 0.3  & 0.290 & 0.065 & 0.297 & 0.058\\
	     & 0.7  & 0.285 & 0.066 & 0.288 & 0.060\\
{\tt pcorr}  & 0.0  & 0.278 & 0.077 & NA      & NA\\
	     & 0.3  & 0.285 & 0.065 & NA      & NA\\
	     & 0.7  & 0.300 & 0.057 & NA      & NA\\
\hline
\end{tabular}\vfill}}\hfill}
\end{table*}
\begin{table*}[t]
\caption{The effect of rounding in the parameters on the fitness evaluation;
$f_{trial}$ is the fitness obtained with \AMORE\ 
for the given values of the parameters. The {\it average value} and
$\sigma$ are based on the 162 models in
Table~\ref{first}. The $\sigma$ is obtained from an
unweighted average for the
fitness of the models involved. The $\widehat{\sigma}$ is the
error as estimated by Ng \cite{Ng98a}.
\label{tbl1}}
\begin{center}
\begin{tabular}{|l|cccccccc|c|}\hline
parameter & log d(pc)& A$_V$ & 
$\log t_{low}$ & $\log t_{hgh}$ & [Z]$_{low}$ & [Z]$_{hgh}$ &
$\alpha$ & $\beta$ & $f_{trial}$\\
\hline\noalign{\smallskip}
\hline
original & 3.906335 & 0\mag00 & 9.90309 & 9.95424 & \minus0.60206 & 0.17609 & 2.35 & 1.0 & 0.44597\\ 
\hline
round-v1 & 3.906 & 0\mag00 & 9.903 & 9.954 & \minus0.60 & 0.18 & 
2.35 & 1.0 & 0.28595\\
round-v2 & 3.906 & 0\mag00 & 9.903 & 9.954 & \minus0.602 & 0.176 & 
2.35 & 1.0 & 0.30812\\
round-v3 & 3.9063 & 0\mag00 & 9.9031 & 9.9542 & \minus0.602 & 0.176 & 
2.35 & 1.0 & 0.42439\\
\hline
\hline
average value & 3.8958 & 0\mag027 & 9.866 & 9.984 & \minus0.554 & 0.244 &
2.358 & 1.574 &\\ 
$\sigma$ & 0.0033 & 0\mag014 & 0.049 &  0.047 &  0.050 & 0.13  &
0.034  & 1.40 &\\
$\widehat{\sigma}$ & 0.012 & 0\mag06 & 0.043 & 0.023 & 0.18 & 0.08 & 0.03
& 1.4 &\\
\hline
\end{tabular}
\end{center}
\end{table*}
Fourthly, we want to explore as large a fraction of the 
parameter space as possible at the first entry in \AMORE. This is done by using
only an one digit accuracy ({\tt nd=1}).
Due to the active re-scaling of the parameter space boundaries 
we do not require a very high precision in our exploration.
A one percent accuracy ({\tt nd=2}) of the parameter space is
sufficient in the subsequent \PIKAIA\ cycles.
\par
In biological terms, the \PIKAIA\ control parameters define the
ecosystem in which our population evolves.
\hfill\break
All computations presented in this paper were performed 
with an executable generated with the {\tt g77} compiler.
This executable was then installed on
various PCs running Red Hat 
Linux\footnote{Red Hat$^{\bigcirc\!\!\!\!{\tt R}}$ 
is a registered trademark of Red Hat Software, Inc.
and Linux$^{\bigcirc\!\!\!\!{\tt R}}$ 
is a registered trademark of Linus Torvalds.} 6.X and 7.X.
The PCs were equipped with Intel Pentium III or Athlon processors
with clock speeds ranging from \mbox{600\to1200~MHz}. 
\par
All tests, unless stated otherwise, use the synthetic population as
described by Ng (\cite{Ng98a}):
\begin{list}{$-$}{\topsep=0pt\parsep=0pt}
\item a metallicity range, spanning Z\muspc=\muspc0.005\to0.030;
\item an age range from 8\to9 Gyr;
\item an initial mass function with a Salpeter slope;
\item an exponential decreasing star formation rate with a characteristic time 
scale of 1~Gyr.
\end{list}
The test population contains $N\!=\!5000$\  stars and is placed at
8~kpc distance. The `observational' limits are set to
\mbox{V$_{lim}$\muspc=\muspc22$^m$} and \mbox{I$_{lim}$\muspc=\muspc21$^m$}. 

\subsection{Description of the tests}
\label{testdesc}
\subsubsection{Test 1: Determining values for {\tt pcross, rcross, rbrood,
pcreep, {\rm and} pcorr}} 
\label{desc_test1}
In the first test, we evaluate the 162 models listed in
Table~\ref{first} in order to study the effect of the \PIKAIA\
parameters {\tt pcross, rcross, rbrood, pcreep, {\rm and} pcorr} on
the convergence and computational effort. The test has as a secondary
objective to provide an understanding of the degeneracy of the
parameter space.\\
All astrophysical parameters to be retrieved are set free, floating
between reasonable minimum and maximum values (see
Table~\ref{amorehrd} for details). \AMORE\ runs for 20
iterations of 
{20}
generations ({\tt ngen=20}) to recover the {\it \`a
priori}\/ known parameters of the synthetic population. 
{The number of iterations and generations 
determine the total length of an evolutionary run: 
$20\!\times\!20=400$ generations. }
Note that the
range of each parameter is set within reasonable limits and not taken
excessively large, because it might lead to the case that no
acceptable parameter setting is found with the standard iteration loop.

\subsubsection{Test 2: Rounding}
The second test deals with the effects of rounding. We vary the
number of significant digits in the input parameters to reveal
\AMORE's sensitivity to rounding. In this case we do not make an
evolutionary run,
because the trial set of parameters (the educated guess at the
start-up of a \PIKAIA\ cycle) is the correct one. For clarity we label
the fitness in this test by $f_{trial}$ instead of $f_A$, the fitness
after a complete evolutionary run.

\subsubsection{Test 3: Fixing parameters at the correct value}
\label{desc_test3}
In the third test we take six models in which one of the parameters is
set fixed at its correct value in order to study the effects on the
convergence. The models chosen were two of high, two of
intermediate and two of low fitness as determined from the first
test. The convergence in this test basically can go two ways: either
the convergence is faster, because less parameters
have to be optimized. Or, due to the fact that \AMORE\ has less
maneuverability in this situation, the convergence is slower.
We adjusted the limits for age and metallicity as given in
Table~\ref{amorehrd} such that \AMORE\ would not try to find solutions
in forbidden regions of parameter space which might severely slow down
convergence due to constant rejection by \AMORE\ of the chosen
parameter values.\\
For example, fixing the Z$_{low}$ parameter at its correct value of
\minus0.60206 means that we have to adjust the lower limit for Z$_{high}$
to \minus0.60206 as well.\\
In the case of fixing the {\it log t$_{low}$} parameter this also
implies that the initial guess has to be adjusted. We set this initial
guess to 10.1.

\subsubsection{Test 4: Fixing parameters at the wrong value}
\label{desc_test4}
In the fourth test we take six models in which one of 
the parameters is set fixed at 1$\sigma$ 
{offset }
(determined
from the first test) from its original value, in order to
study its effect on the `second best' setting of the remaining parameters.
Normally one would expect a fitness \mbox{$f\!>\!{1\over3}$}.
In this case, however, 
\mbox{$F\!<\!F_P^2+F_\chi^2 = 1^2+(1+1)^2 = 5$} and the
associated fitness constraint drops to \mbox{$f\!>\!{1\over6}$}.
{
However, this assessment ignores the fact that, when a parameter is
offset from its optimum value, the number of matched points will
decrease and $F_P$ increases. Using Eq.~(\ref{eq average sigma})
one has for a good fit $F\!=\!2$. On average the offset 
per parameter $k$\/ from the optimum value is 
$\sqrt{1\over4}\,\sigma_k\!=\!{1\over2}\,\sigma_k$, at best 
the offset is $0\,\sigma_k$, and in the worst case this is 
$\sqrt{2}\,\sigma_k$.
So with one parameter $k$\/ put at $1\,\sigma_k$ offset we 
distinguish the three possibilities
\begin{tabular}{llllll}
1&at best   &$F\!=\!(1+2)$                     & $=3 $  & $\rightarrow $ 
&$f\!=\!{1\over4}$\\
2&on average&$F\!=\!(1+{1\over2})^2+{7\over4}$ & $=4 $  & $\rightarrow $ 
&$f\!=\!{1\over5}$\\
3&at worst  &$F\!=\!(1+\sqrt{2})^2$            & $=5.8$ & $\rightarrow $ 
&$f\!=\!{1\over6.8}$ 
\end{tabular}
Note that the worst case limit is in agreement
with the results presented in Table~\ref{fixwrong_tbl}.
}
\\
The effect of the \mbox{1\muspc$\sigma$} offset of one of the parameters 
will partly be canceled by forcing other parameters away
from the optimum value. For example, the effect of an increased 
extinction can be masked partially by generating a bluer
stellar population with a lower metallicity and a younger age.
The effect will be such that the fitness will not
be around \mbox{$f\!\simeq\!{1\over{6.8}}$}, but 
somewhere in the range \mbox{${1\over{6.8}}\!<\!f\!<\!{1\over3}$}.\\
We fixed the parameters both at one sigma above and one sigma below
the original value, because 
the evolutionary effects do not have to be
symmetric. The only exception is the extinction, which we only fix at
one sigma above the original value of A$_V$ = 0.0\\
Again we adjusted the limits for the upper and lower limit for age and
metallicity.

\section{Results}
\label{Results}
\begin{figure*}[t]
\resizebox{18.0cm}{!}{\includegraphics{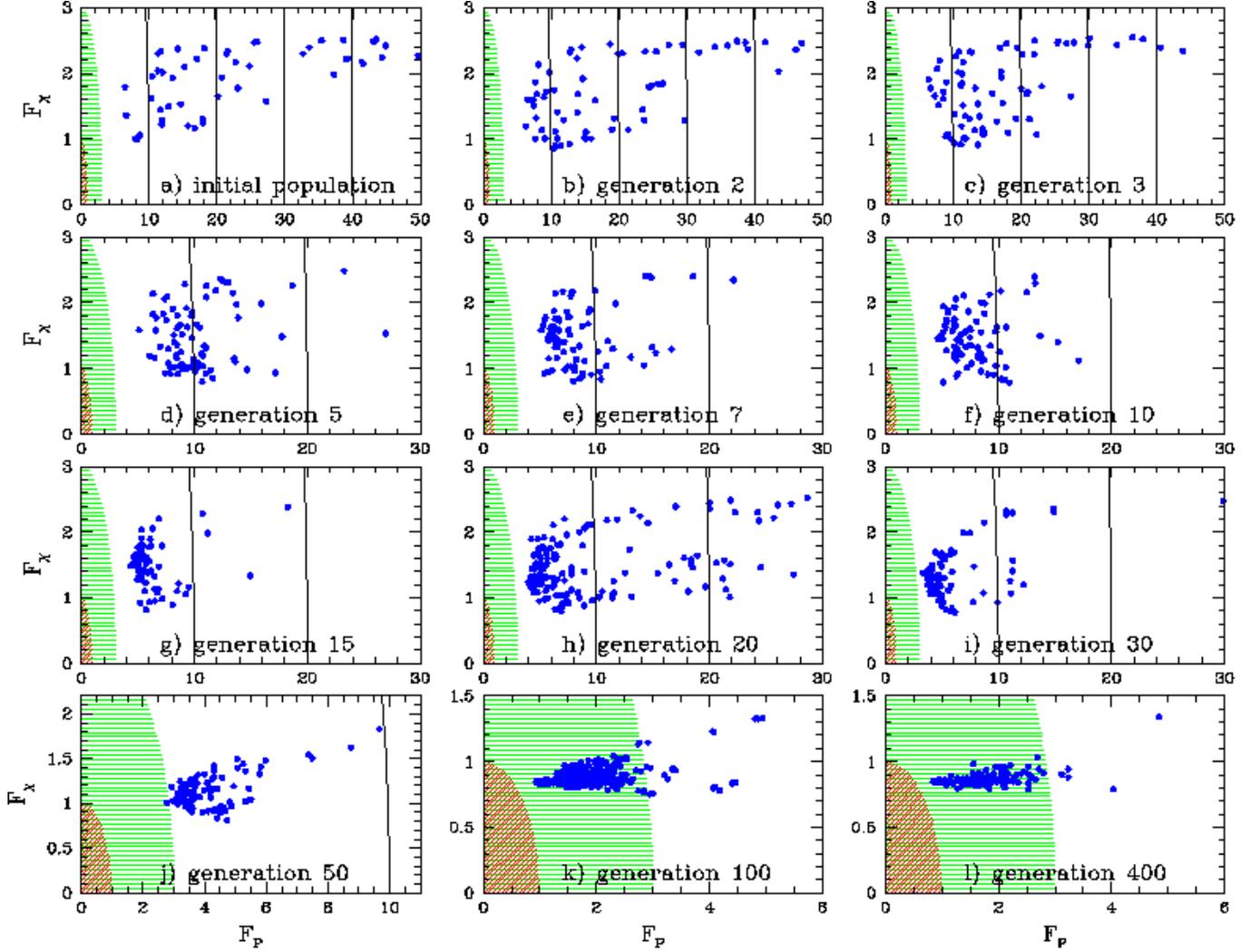}}
\caption{\label{generationfig}
Conception diagram of the 
evolution of the genetic population of model \ref{first}-40 during 
the optimization process displayed in Fig.~\ref{evolvefig}.
Frame {\bf a} shows the initial population and the frames 
{\bf b}\to{\bf l} show the population after several generations
up to generation=400. The 
{ outer }
shaded region indicates solutions for which
the difference between the CMDs from the `observed' and synthetic
population is on average 
{ between 1\to}3\muspc$\sigma$.
{ The inner shaded regions marks the region with 
solutions for which the difference between the `observed' and synthetic
CMDs are less than 1\muspc$\sigma$. Such solutions are close to 
perfect matches between the `observed' and synthetic data 
and are considered to belong to a group of solutions 
for which one may say ``too good to be true''.}
The solid lines indicate
the 10\muspc$\sigma$, 20\muspc$\sigma$, 30\muspc$\sigma$ 
and 40\muspc$\sigma$ contours}
\end{figure*}

\begin{figure*}
\begin{center}
\resizebox{4.25cm}{!}{\includegraphics{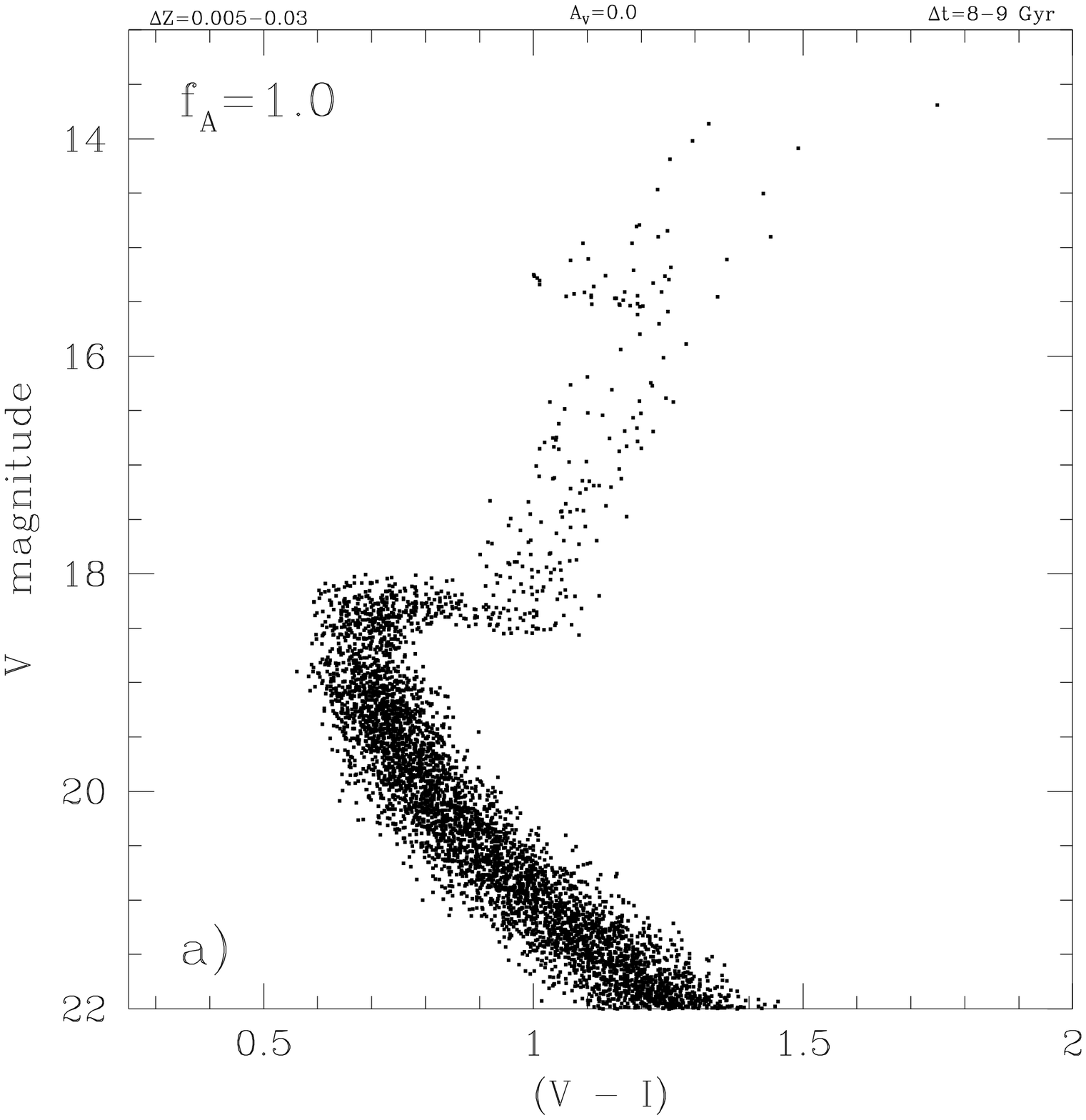}}\quad
\resizebox{4.25cm}{!}{\includegraphics{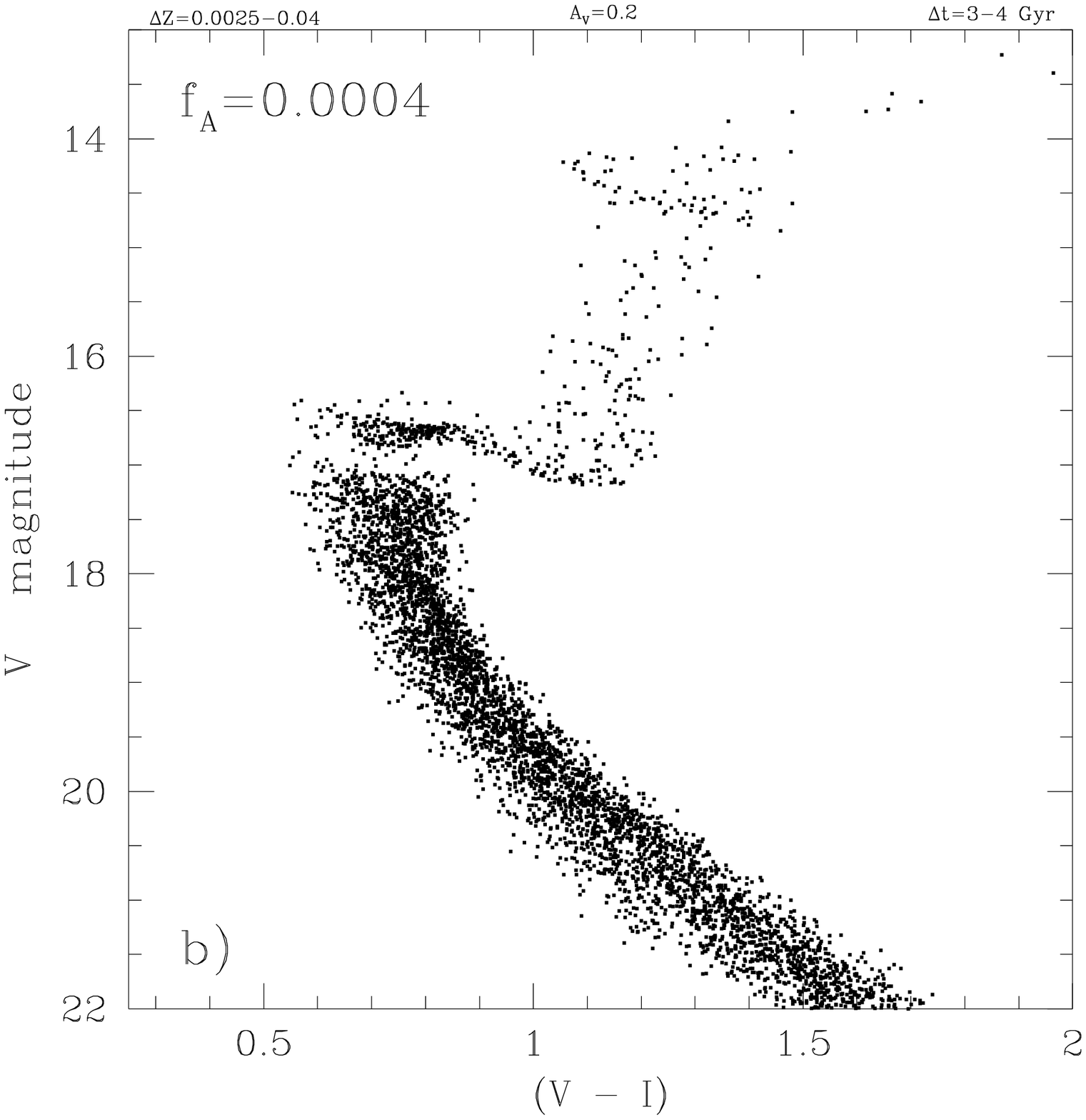}}
\end{center}
\begin{center}
\resizebox{4.25cm}{!}{\includegraphics{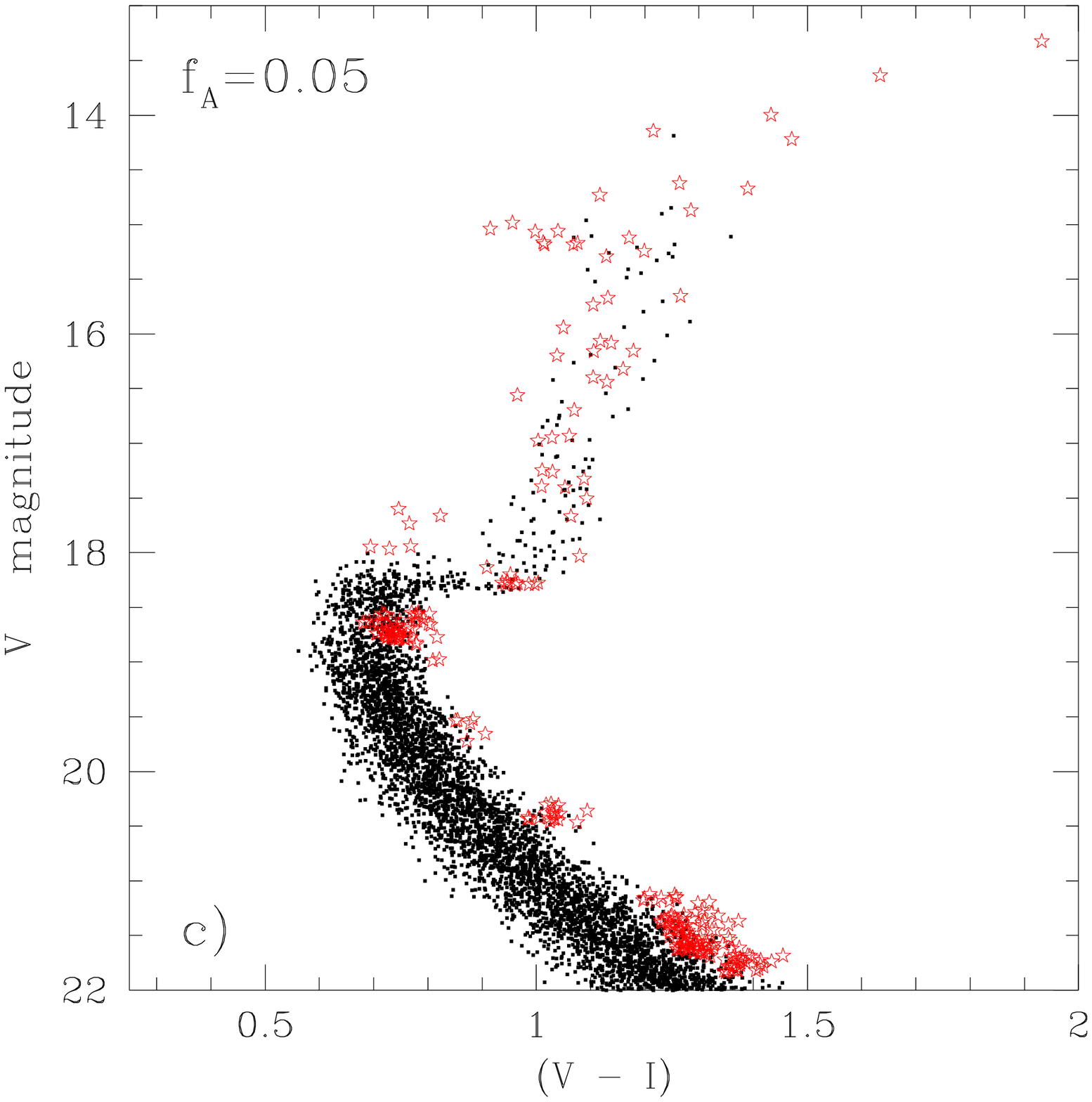}}\quad
\resizebox{4.25cm}{!}{\includegraphics{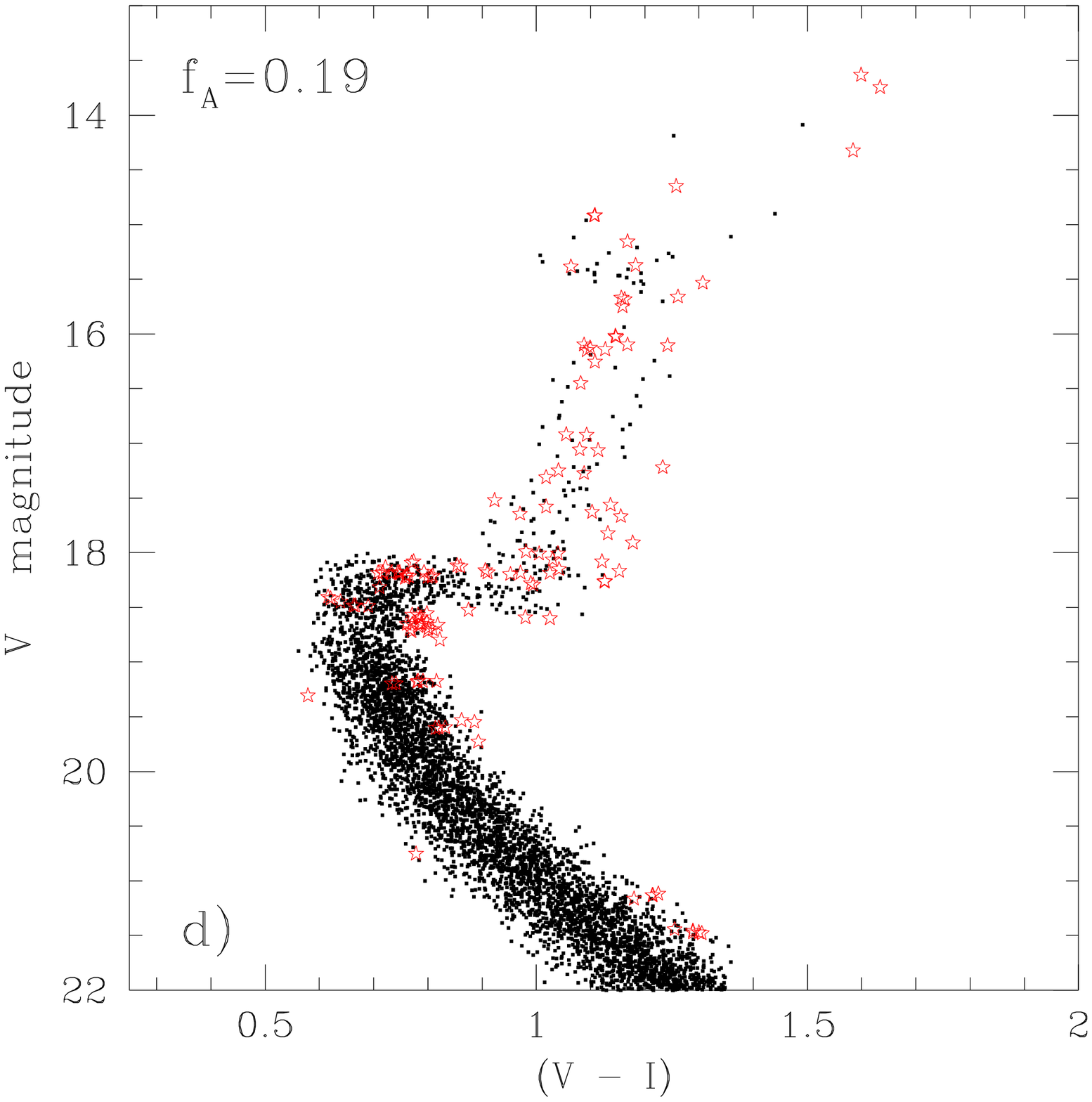}}\quad
\resizebox{4.25cm}{!}{\includegraphics{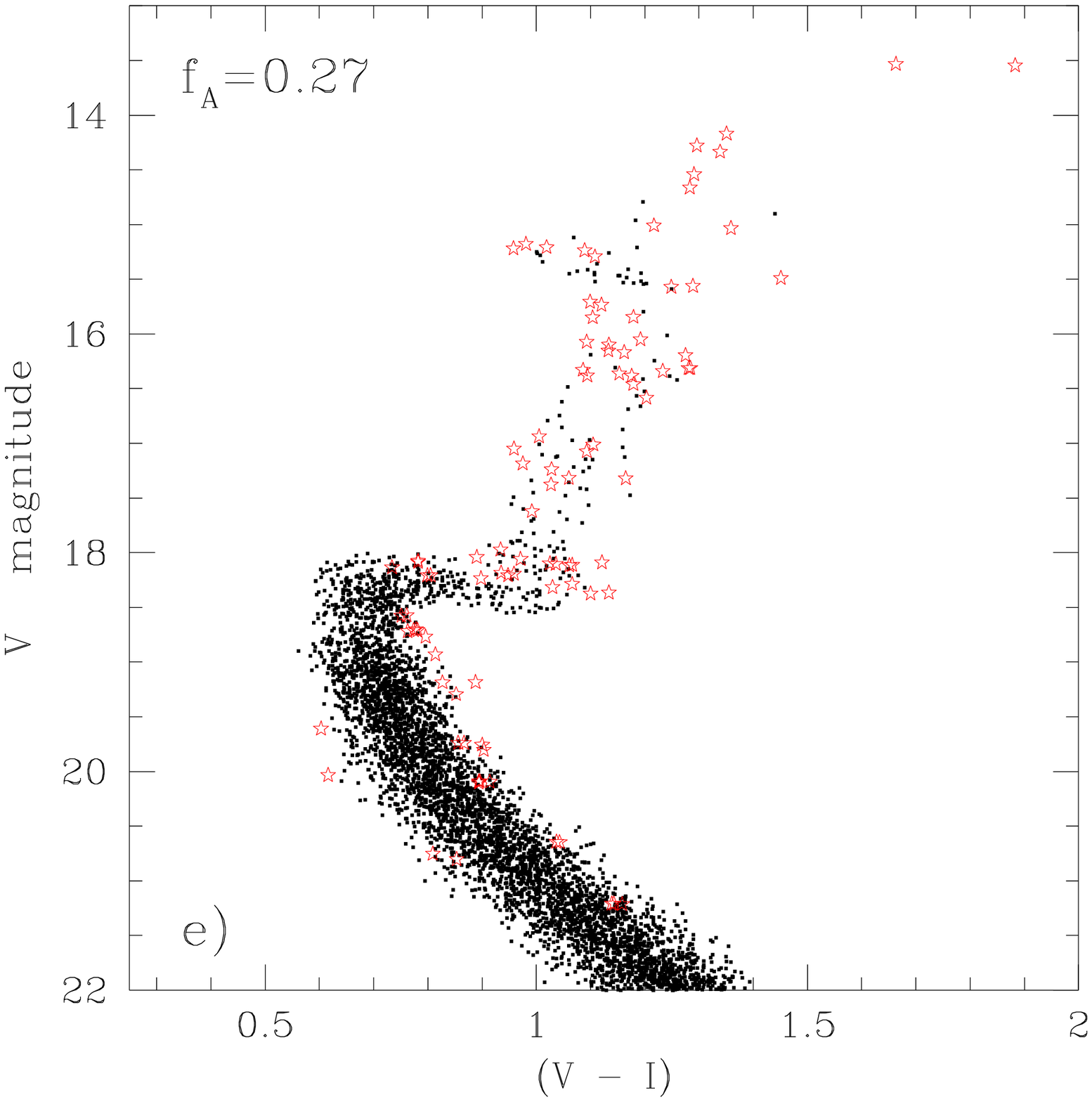}}\quad
\resizebox{4.25cm}{!}{\includegraphics{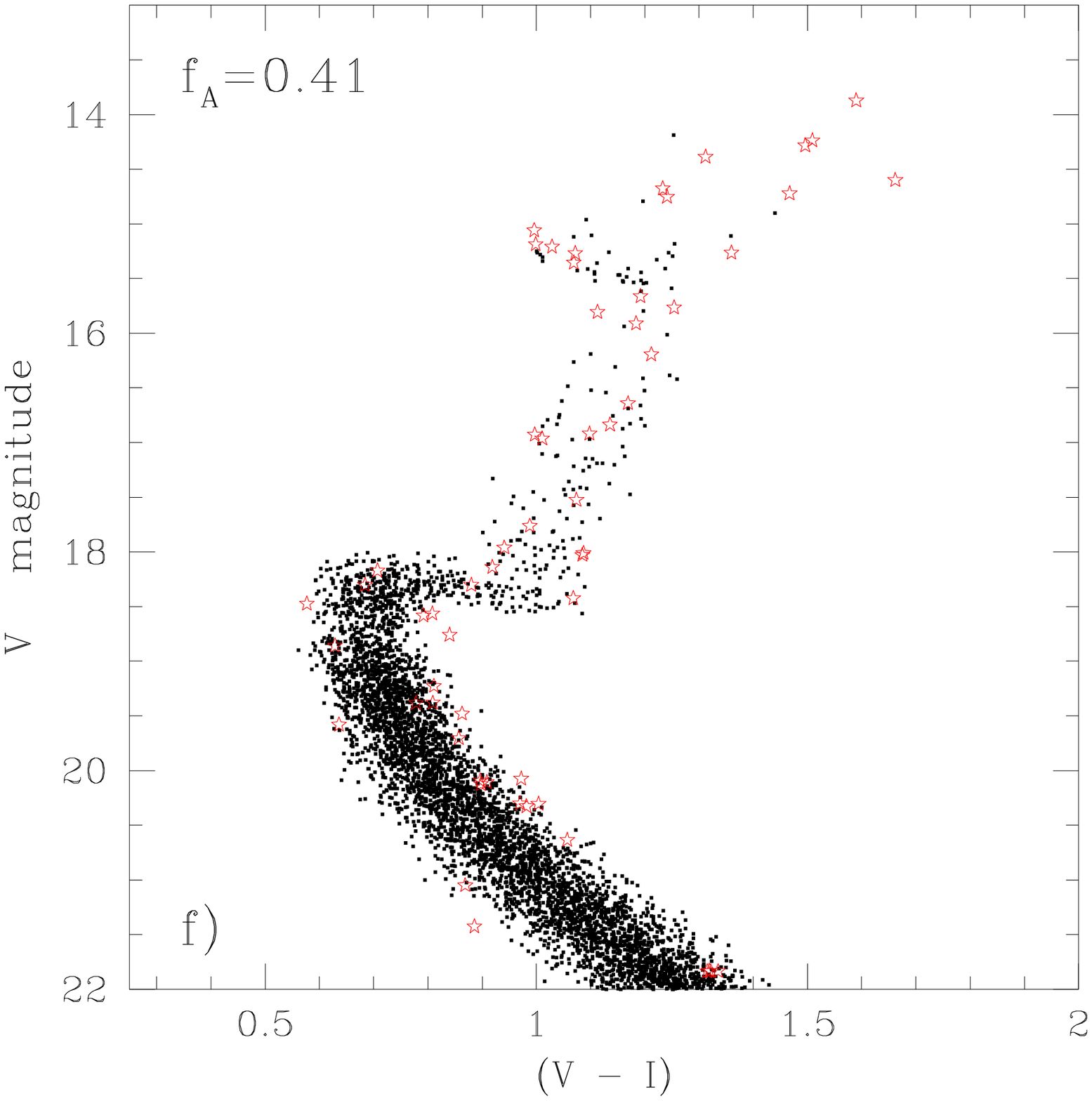}}
\end{center}
\begin{center}
\resizebox{4.25cm}{!}{\includegraphics{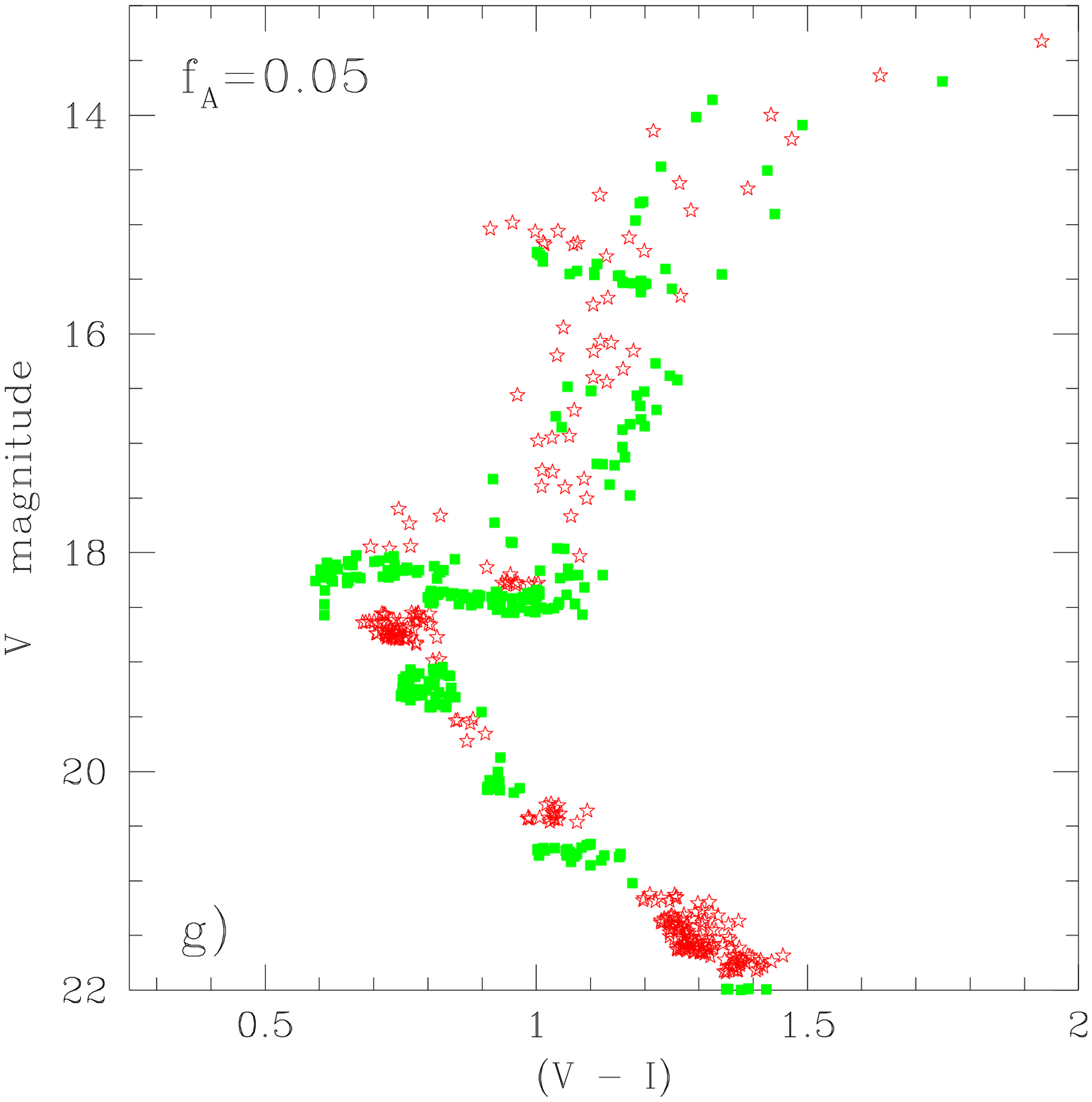}}\quad
\resizebox{4.25cm}{!}{\includegraphics{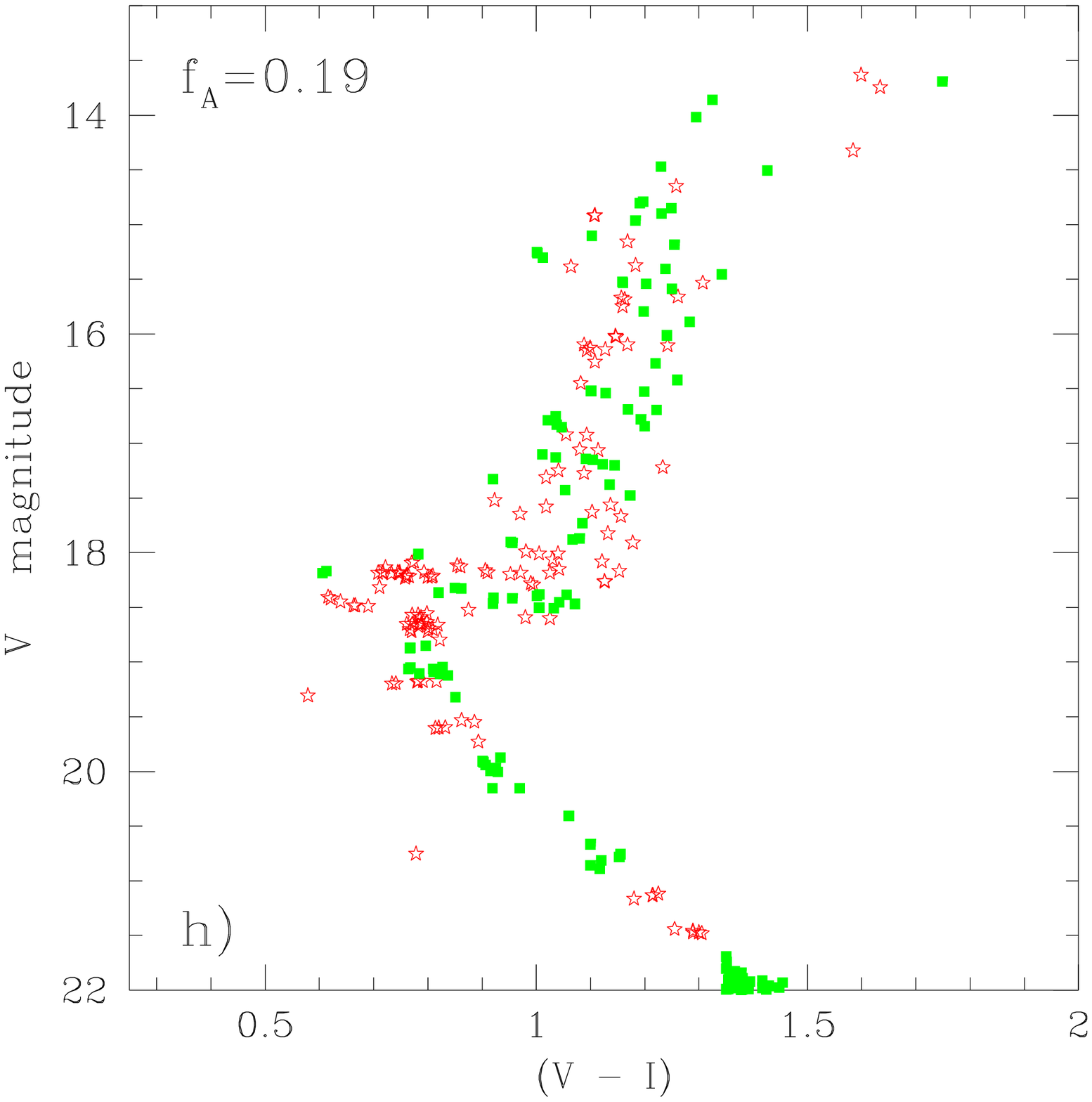}}\quad
\resizebox{4.25cm}{!}{\includegraphics{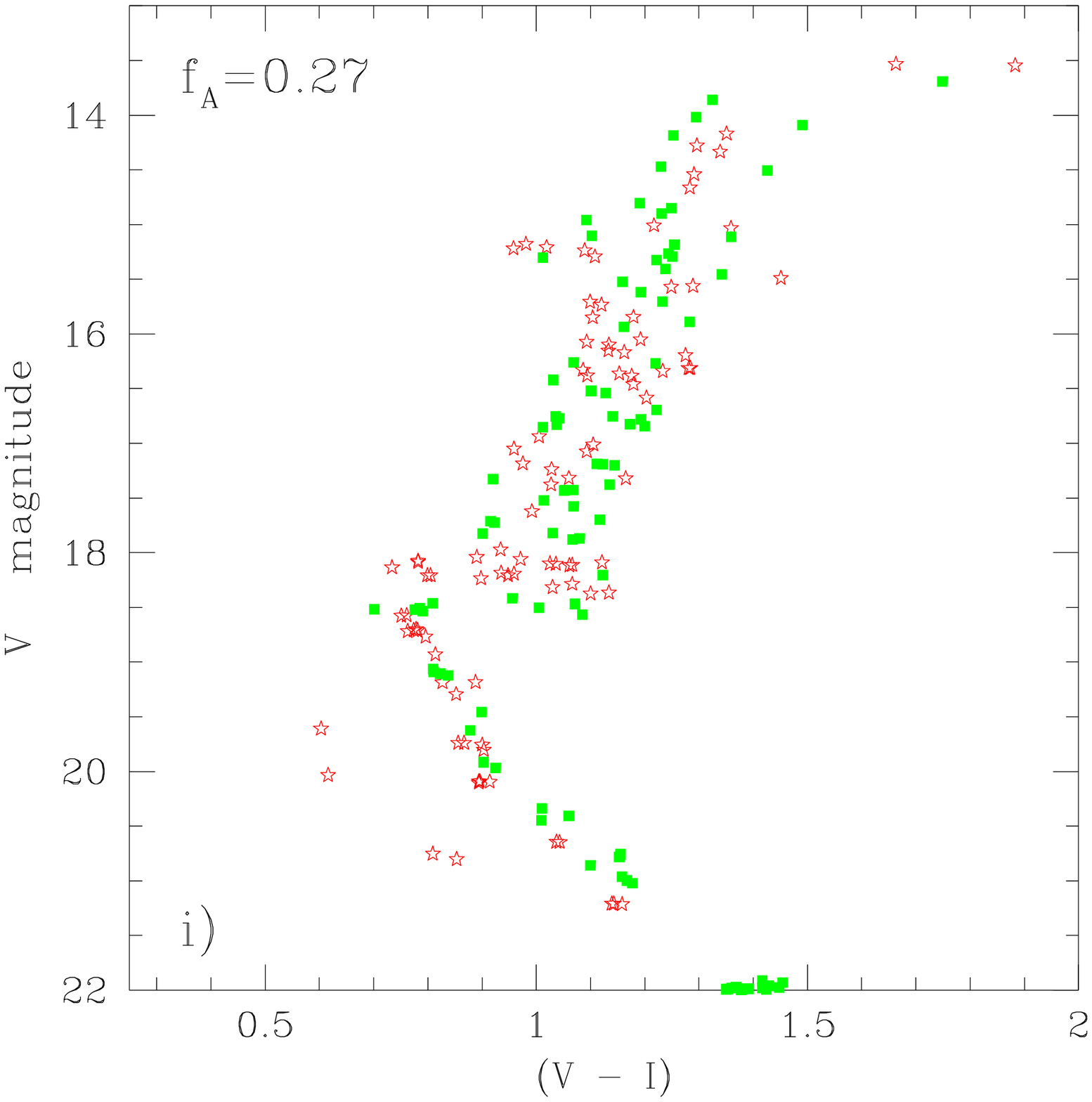}}\quad
\resizebox{4.25cm}{!}{\includegraphics{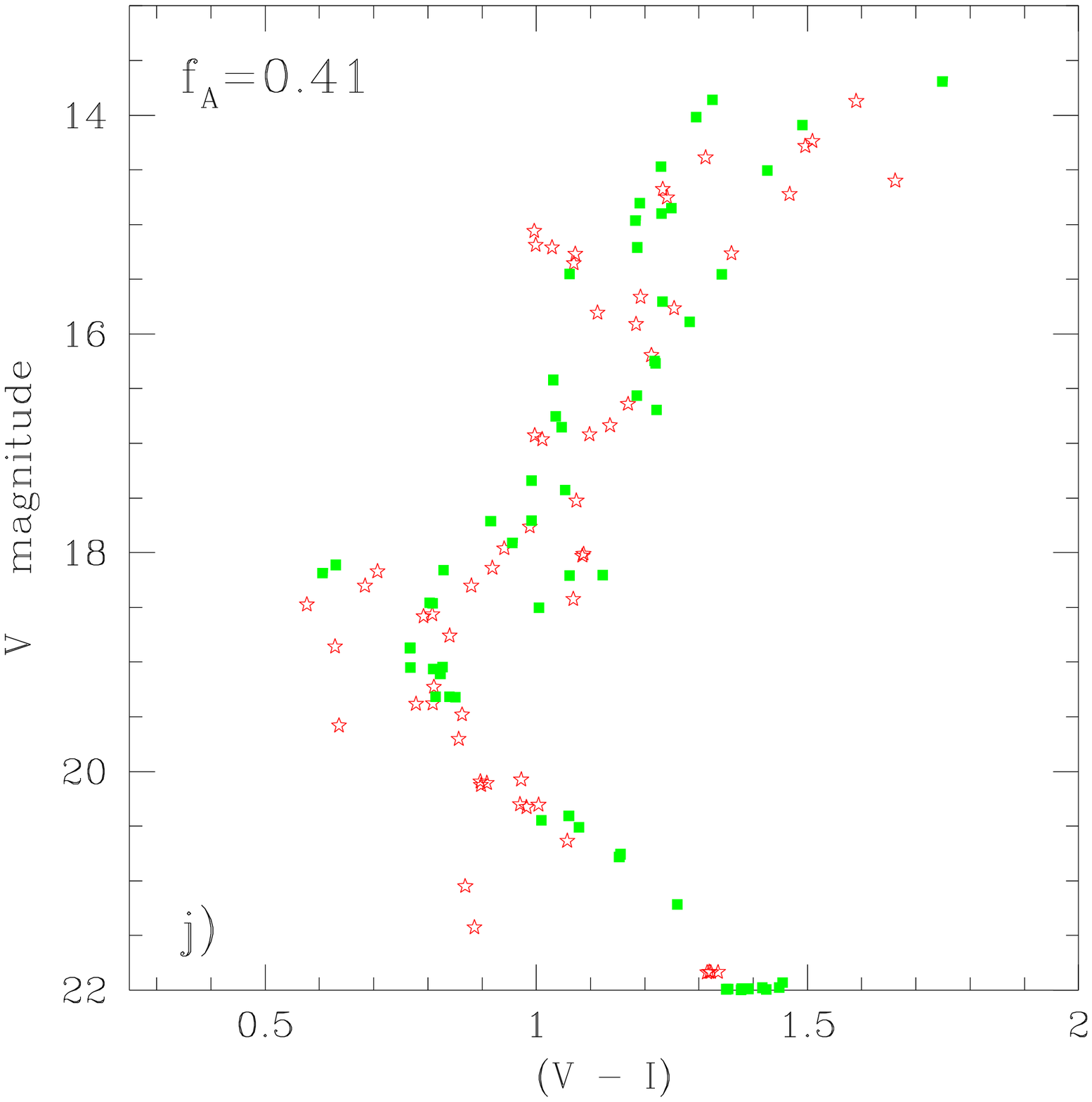}}
\end{center}
\caption{\label{hrdevolve}
Genetic evolution of the colour-magnitude diagram (CMD)
from the first test population. 
Panel {\bf a} displays the original population to be matched.
The physical parameters for this population are described
Table~\ref{amorehrd} and Sect.~\ref{setup}. The CMD of the initial trial
population is shown in panel {\bf b}. Panels {\bf c\to{e}} display the
resulting CMDs obtained with setup \ref{first}-40 for different
fitnesses (see Sect~\ref{fitness}). The fitnesses $f\!=\!0.05$, $f\!=\!0.19$
and $f\!=\!0.27$ are respectively reached after 20, 60 and 80 generations.
The dots in the panels {\bf b\to{e}} are used for each matching point, 
while the red open stars {\color{red}$\bigstar$} in panels {\bf c \to{j}}
are the points in the simulation which have no counterpart in the
original CMD. Panel {\bf f} displays the fitness $f\!=\!0.41$ 
as obtained after 
341 generations. Note that the CMDs of panels {\bf a} and {\bf f}
as well as {\bf c\to{e}} are visually almost indistinguishable.
Panels {\bf g\to{j}} displays the residuals between the simulated and
the original CMD ({\bf a}; green solid squares
{\color{green}$\blacksquare$}) are those points in the original which
have no counterpart in the simulation}
\end{figure*}

{\AMORE\ has been tested for a wide range of setups.
The results in Table~\ref{first} indicate that \AMORE\ 
give both acceptable and less acceptable solutions.  
They are displayed in Figs.~\ref{generationfinfig} \& 
\ref{evolveamorefig}. \par
To better understand what goes on during the genetic evolution
we display the results from model \mbox{\ref{first}-40}.  
Figure~\ref{generationfig} displays an example of the
evolution of the merit function $F$ for a number of generations.
It shows how the initially dispersed individuals gradually find
their way, start to cluster together around generation 10,
and penetrate the region with acceptable solutions after 
about 50 generations. After 100 generations the improvements
become marginal for this model.   
\\
Figure~\ref{hrdevolve} displays for the same model \mbox{\ref{first}-40}
the phenotypical changes of the CMD for several fitnesses during the 
genetic evolution. The various panels show that the synthetic 
CMD resembles better and better the `observed' CMD when the 
fitness improves. Note that at fitness \mbox{$f$\muspc=\muspc0.05} 
one already gets for the eye appealing solutions. 
\\
Figure~\ref{evolvefig} shows the improvements of the 
astrophysical parameters as a function of increasing fitness
for the models \mbox{\ref{first}-40} and \mbox{\ref{first}-51}.
The panels for distance and extinction show that the 
distance is systematically underestimated, while the 
extinction is overestimated. But in general one notices that
the astrophysical parameters obtained from model 
\mbox{\ref{first}-40} get quite close to the parameters 
of the CMD to be matched. }


\subsection{Test 1: Parameter values and degeneracy}
\subsubsection{Degeneracy of the parameter space}
\label{res_test1}
Table~\ref{first} is displayed in
Fig.~\ref{generationfinfig}. The clustering in the figure provides
an indication that a degeneracy of the parameter space is present near 
$f\!>\!0.25$ (i.e. $F\!<\!3$, {see eq.~(\ref{eq fitness})}). 
Without a major computational effort
it will be difficult to obtain a significant improvement of the 
parameters once $f\!>\!0.25$.
However, $F\!<\!3$ indicates a region in the ($F_\chi,F_P$)-plane
for which the 
{systematic offset of the individual parameters 
from its true value }
are on average less than { $\sqrt{F/n}\,\sigma_k\!=\!0.6\,\sigma_k$, 
see Sect.~\ref{contract} for details}. 
In practice it turns out that 
a strong correlation between three of the eight parameters 
has the culprit; at least two of them have to change simultaneously 
in the proper direction in order to improve the fitness
(see also Sect.~\ref{degeneracy}).
They have an 
{ average offset of
$\sim\!\sqrt{F/3}\,\sigma_k\!=\!1\,\sigma_k$}, 
while for the remaining parameters  
this is $<\!<1\,\sigma_k$.
\par
\begin{figure*}[t]
\resizebox{18.0cm}{!}{\includegraphics{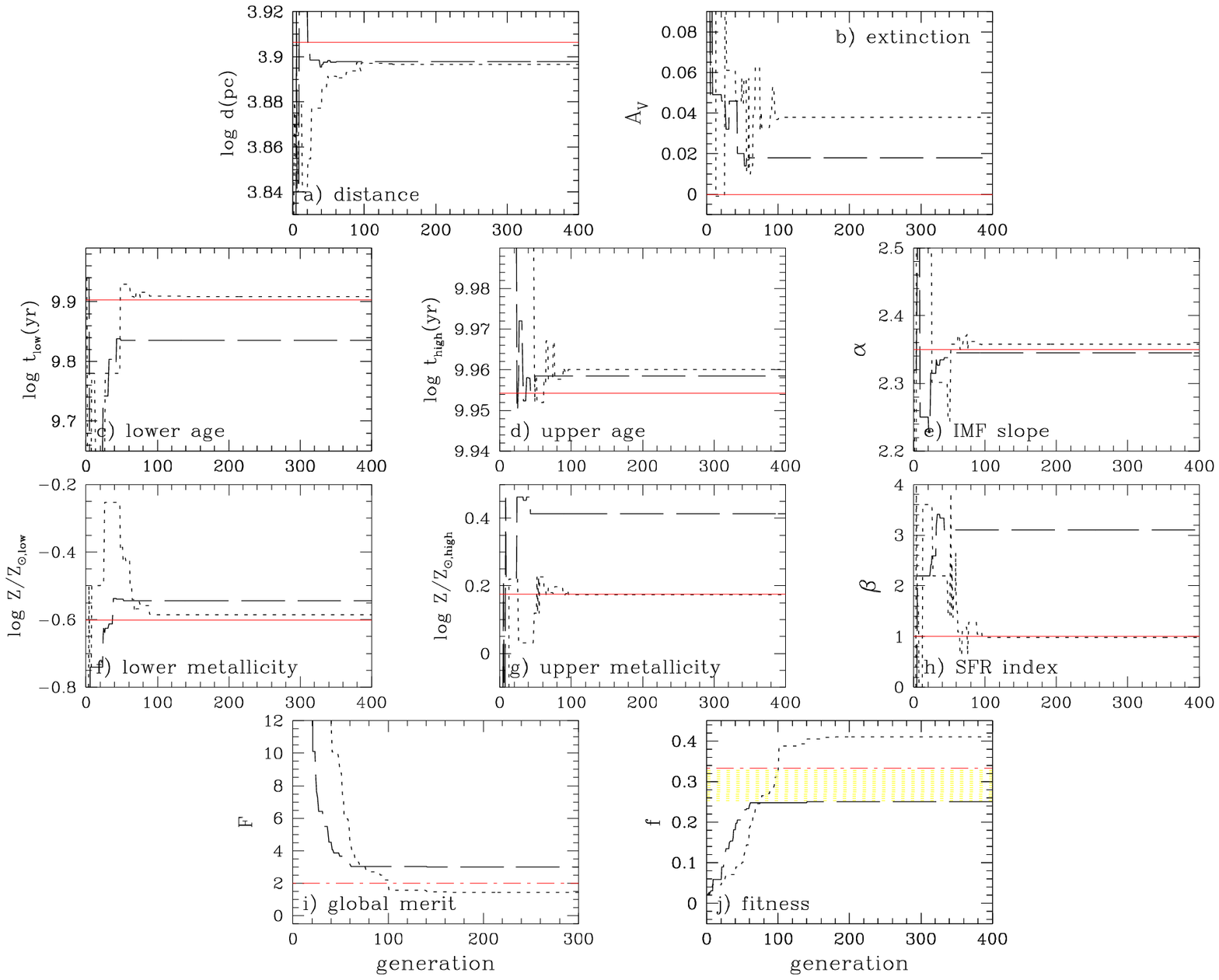}}
\caption{\label{evolvefig}
Panels {\bf a}\to{\bf h} display the 
convergence curves for the parameters of models 
\ref{first}-40 (dotted line; fitness $f=0.41$) 
and \mbox{\ref{first}-51} (long dashed line; fitness $f=0.25$).
The solid line in the frames {\bf a\to{f}} refers
to the value adopted for the original population.
The long, dot dashed line in frames {\bf i\muspc\&\muspc{j}}
shows the threshold values to be crossed for acceptable solutions,
i.e. $F\!<\!2$ and ${1\over3}\!<\!f\!<\!1$.
The short dashed area in frame {\bf j} marks the region 
where degeneracy of the parameter space becomes noticeable 
(see Sects.~\ref{degeneracy} for details)
}
\end{figure*}

In addition, Fig.~\ref{evolveamorefig} displays the retrieved
parameters for all models
as a function of fitness. Note, that \AMORE\ systematically
underestimates the distance of the test population. On the other hand,
the effect of this underestimation is in its turn partially canceled by  
overestimating the extinction, the upper 
age limit and the slope of the power-law IMF slightly
{(see also Figure~\ref{evolvefig})}.
Another clue we get from Fig.~\ref{evolveamorefig} is that the slope
of the SFR $\beta$ is very poorly constrained.

\subsubsection{Determining values for {\tt pcross, rcross, rbrood,
pcreep, {\rm and} pcorr}}
\label{detervalues}
Looking at Table~\ref{first}, a result that is immediately obvious is
that {\tt pcorr = 0.0} has a strong
stabilizing effect on the simulation. Its effects overwhelm the
effects of the {\tt pcreep} parameter and lock $f_A$ at a certain
value, which may be quite good (e.g. models 10, 11 and 12) or quite
poor (e.g. models 28, 29 and 30). Setting {\tt pcorr} $>$ 0 is thus
preferred to avoid getting locked in a low value for $f_A$.\\
In order to compensate for this strong stabilizing effect, we also
evaluate in Table~\ref{avgfit}
the average fitness of the models when we exclude all models which
have {\tt pcorr} = 0.0.\\
As expected, the {\tt rbrood} parameter has a strong influence on the
amount of computational time needed. Although the models 
with high values of {\tt rbrood} are somewhat better than
models with low values, this effect is only marginal. Considering that a
high value of {\tt rbrood} lessens the genetic variation in the gene
pool while increasing the computational time needed for a run with
several factors, it is desirable to have a low value of {\tt rbrood}.\\
\\
The different parameters are not independent, as can be seen from
Table~\ref{first} and Table~\ref{avgfit}.
Simply taking the best options in Table~\ref{avgfit} yields
model 134 for the case in which {\tt pcorr=0.0} has not been corrected
for, a reasonable, but not an exceptionally good model.

\subsection{Test 2: Rounding}
\label{res_test2}
Table~\ref{tbl1} shows the effect of the accuracy of the retrieved 
values of the parameters on the evaluation of the fitness. If one
applies a rounding to one or two significant digits it is not possible
to reproduce the expected fitness, i.e. the expected fitness drops
from $f_{trial}\!=\!0.44$ to $f_{trial}\!=\!0.28$. 
A better agreement can be obtained by reporting the values of 
the parameters with the addition of one or more 
(apparently insignificant) digit(s). 
A renewed search (Table~\ref{tbl1}; round-v3)
circumvents the local optimum at $f_{trial}\!=\!0.28$ and
results in a near global fitness of $f_{trial}\!=\!0.42$,
which is close enough to the value expected.\\
The {\it true $\sigma$} line in the table shows that both the
Z$_{high}$ and the $\beta$ parameter are the weak links in the overall
parameter estimation (see also Fig~\ref{evolveamorefig}).
\begin{table}
\begin{center}
\caption{Fitness statistics when fixing one parameter at its correct
value. Averaged fitness values $\overline{f_A}$
and their associated standard deviation $\sigma_{n-1}$
are obtained from simulations with the 
setup parameters from models 9, 14, 22, 34, 40 and 52. See
Table~\ref{fixgood_tbl} for additional details
\label{statgood_tbl}}
\begin{tabular}{|l|cc|}
\hline
parameter & $\overline{f_A}$ & $\sigma_{n-1}$ \\
\hline
\hline
log d          &  0.299  &  0.081\\
A$_V$          &  0.274  &  0.081\\
$\log t_{low}$  &  0.332  &  0.021\\
$\log t_{high}$ &  0.304  &  0.056\\
Z$_{low}$      &  0.269  &  0.036\\
Z$_{high}$     &  0.361  &  0.036\\
$\alpha$       &  0.305  &  0.066\\
$\beta$        &  0.312  &  0.076\\
\hline
\end{tabular}
\end{center}
\end{table}

\subsection{Test 3: Fixing parameters at the correct value}
\label{results_test3}
The results of fixing parameters at the correct value are listed in
Table~\ref{fixgood_tbl} and an example of the diagnostics is listed in
Table~\ref{diag_test3}. Details of the individual setups 
for these tests are given below.
In general, the results of the tests 
for which one of the parameters was set 
to the correct value were slightly better than the results 
for the models for which all parameters are set free,
see Tables~\ref{tbl1} \& \ref{statgood_tbl} for additional details.
This behaviour 
is due to the fact that by forcing one parameter to a fixed value 
the evolutionary path changes.
The models were selected from the results 
with low and intermediate fitness given in Table~\ref{first}. 
\begin{table*}
\caption{Description of the diagnostic statistics for
model \ref{first}-40 when fixing one parameter at its correct value, 
see Sects.~\ref{results_test3},\ref{fitness} and \ref{unstable} for
additional details}
\label{diag_test3}
\begin{center}
\begin{tabular}{|l||ccc|ccc||c|}
\hline
model & $N_{O,not}$ & $N_{S,not}$ & $N_{match}$ & $F_\chi$ & $F_P$ 
& $F$ & $f_A$ \\ 
\hline
\hline
ideal & 0 & 0 & 5000 & 0.000 & 0.000 & 0.000 & 1.000 \cr
free  & 60 &  60  & 4940 & 0.847 & 0.849 & 1.438 & 0.410 \cr
\hline
\hline
fixed: log d (pc) & 155 & 155 & 4845 & 0.903 & 2.192 & 5.621 & 0.151 \cr
fixed: A$_V$      & 103 & 103 & 4897 & 0.840 & 1.457 & 2.827 & 0.261 \cr
fixed: $\log t_{low}$(yr) & 76 & 76 & 4924 & 0.817 & 1.075 & 1.823 & 0.354 \cr
fixed: $\log t_{hgh}$(yr) & 104 & 104 & 4896 & 0.902 & 1.471 & 2.978 & 0.251\cr
fixed: $[{\rm Z}]_{low}$ & 94 & 94 & 4906 & 0.836 & 1.329 & 2.466 & 0.289 \cr
fixed: $[{\rm Z}]_{hgh}$ & 76 & 76 & 4924 & 0.815 & 1.075 & 1.820 & 0.355\cr
fixed: $\alpha$ & 74 & 74 & 4926 & 0.834 & 1.047 & 1.799 & 0.357 \cr
fixed: $\beta$  & 80 & 80 & 4920 & 0.727 & 1.131 & 1.808 & 0.356 \cr
\hline
\end{tabular}
\end{center}
\end{table*}

\subsubsection{Fixed distance and extinction}
\label{res_fixageAv}
Distance and extinction are 
negatively correlated. When
both parameters are set free, a certain degree of degeneracy is to be
expected. Fixing one of the parameters at its correct value can break
this degeneracy.
The result depends strongly on the evolutionary path of the other
parameters.\\
The lower value for the
average fitness $\overline{f_A}=0.299\muspc\pm\muspc0.081$,
when fixing the distance at the correct value, 
is caused by the presence of 
one outlier (see Table~\ref{fixgood_tbl}),
which is caused by the age-metallicity degeneracy. 
Excluding this value results in
an average fitness of $\overline{f_A}=0.328\muspc\pm\muspc0.039$. 
In general: the extinction
can be reliably retrieved when fixing the distance.\\
When considering a fixed extinction, the results show a strong variation in
both age and metallicity. It should also be noted that the average
{ $^{10}\log$\muspc(distance (pc)) }
retrieved is only \mbox{3.8953\muspc$\pm$\muspc0.0066}. 
This is more than one sigma away from the optimum value for the distance (see
Table~\ref{tbl1}). This is an indication that retrieval of the 
distance by fixing the extinction is hampered by 
the age-metallicity degeneracy. Therefore, the distance cannot be
reliably retrieved when fixing the extinction to its correct value. 

\subsubsection{Fixed age and metallicity}
\label{Fixed age and metallicity}
{Fixing one of the age limits results in values for both the age
and metallicity which are close to the input values. }
This is due to the (partial) breaking of the
age-metallicity degeneracy. The distance-extinction
degeneracy remains. 
{ The results also suggest that the age-metallicity
degeneracy has a stronger impact on the fitness than the
distance-extinction degeneracy.\\ 
Fixing the upper metallicity limit to its 
correct value shows that the values for age and metallicity 
come closer to their original, input values. }
This is quite in 
contrast with the results obtained from fixing the 
lower metallicity to its correct value.
Table~\ref{fixgood_tbl} shows that both the high metallicity
limit and the slope of the exponential SFR are not well
constrained. This behaviour can be accounted to the 
implicit shape of the linear age-metallicity relation
adopted in the \HRDGST. The number of high metallicity 
stars is smaller than the number of low metallicity
stars due to the adopted, exponentially decreasing ($\beta=1$),
star formation rate. The consequence is that the high metallicity 
limit can be better determined when both the low \& high
limits are determined in union. 
\begin{table}
\begin{center}
\caption{Fitness statistics when fixing one parameter
$\pm\muspc1\sigma$ from its original value. Averaged
fitness values $\overline{f_A}$ and their associated 
standard deviation $\sigma_{n-1}$
are obtained from simulations with the setup parameters
from models 9, 14, 22, 34, 40 and 52. See Table~\ref{fixgood_tbl} for
additional details}
\label{statwrong_tbl}
\begin{tabular}{|lc|cc|}
\hline
parameter & offset & $\overline{f_A}$ & $\sigma_{n-1}$\\
\hline
\hline
log d          & $-1\sigma$ &  0.291  &  0.048\\
               & $+1\sigma$   &  0.280  &  0.066\\
A$_V$             & $+1\sigma$   &  0.284  &  0.020\\
$\log t_{low}$  & $-1\sigma$ &  0.258  &  0.044\\
               & $+1\sigma$   &  0.279  &  0.012\\
$\log t_{high}$ & $-1\sigma$ &  0.229  &  0.022\\
               & $+1\sigma$   &  0.207  &  0.014\\
Z$_{low}$      & $-1\sigma$ &  0.254  &  0.066\\
               & $+1\sigma$   &  0.315  &  0.027\\
Z$_{high}$     & $-1\sigma$ &  0.270  &  0.054\\
               & $+1\sigma$   &  0.283  &  0.018\\
$\alpha$       & $-1\sigma$ &  0.266  &  0.025\\
               & $+1\sigma$   &  0.273  &  0.027\\
$\beta$        & $-1\sigma$ &  0.232  &  0.071\\
               & $+1\sigma$   &  0.276  &  0.011\\
\hline
\end{tabular}
\end{center}
\end{table}

\subsubsection{Fixed slope for the power-law IMF}
Fixing the slope $\alpha$ of the power-law IMF
to its correct value
ensures stability in the magnitude direction of the
CMD. This implies that the degeneracy in the colour dependent
parameters, like age, metallicity, and partly the 
star formation rate, becomes more apparent.
Although the overall results are quite good, 
only model \ref{fixgood_tbl}-14 is significantly affected by this 
degeneracy. If we leave out model \ref{fixgood_tbl}-14 
from the statistics the average
fitness becomes $\overline{f_A}=0.331\pm0.020$.

\subsubsection{Fixed slope for the exponential SFR}
The exponential star formation rate parameter $\beta$
is tied to the age \& metallicity range, see also 
Sect.~\ref{Fixed age and metallicity}.
Fixing the parameter $\beta$ better constrains in particular
the upper metallicity limit. However, it does not 
avoid that the genetic evolution enters into 
a local age-metallicity gap, see Table~\ref{fixgood_tbl}.
Excluding model \mbox{\ref{fixgood_tbl}-34} improves the average fitness
in Table~\ref{statgood_tbl} to $\overline{f_A}=0.342 \pm 0.029$.

\subsection{Test 4: Fixing parameters at the wrong value}
\label{results_test4}
The results of fixing the parameters at a $\pm\muspc1\sigma$
offset from its original value are listed in Table~\ref{fixwrong_tbl}. 
An example of the diagnostics of these tests are given in
Table~\ref{tbl3}. Details of individual setups are given below.

\subsubsection{Erroneous distance}
\label{wrong_distance}
A wrong assumption about the distance gives in a vertical
shift in the CMD (all other parameters give a diagonal shift) and
cannot be masked out
through a correlated change of any of the other parameters.
This results in quite a wide range in the values of the other
parameters, except the value of $\alpha$.
{The unexpected result is that irrespective if 
the distance is too short or too far: the power-law IMF slope 
flattens!\\
If the distance is overestimated there are more synthetic stars
present at fainter magnitudes. To get relatively more synthetic 
stars at brighter magnitudes one needs to flatten the power-law IMF slope. 
\\
If the distance is underestimated then more synthetic stars are
present at brighter magnitudes. One expects that 
a steeper power-law IMF slope is required as compensation.
This is not always true, see Table~\ref{fixwrong_tbl}. 
Stars pop up at the lower end of the main sequence.
They are taken away from the stars located at brighter and brighter 
magnitudes. One therefore requires also in this case a flatter IMF slope.  
}
\par
A flatter slope of the 
power-law IMF can be a hint that the distance of the
stellar aggregate is wrong. Or it might be a hint that the zero point 
of the adopted synthetic photometric system is different from the
actual photometric system used.\par
Recognizing that the slope is indeed flatter than
the majority of the other cases outlined in Table~\ref{tbl1}
one may start to explore the assumption that the 
distance is wrong: release the constraint 
during the next exploration. 
\par
The sensitivity to the distance implies that \AMORE\  can be used to
determine the distance to a stellar aggregate quite reliably. A
bonus is that due to an initially wrongly assumed distance the
extinction is in most cases better constrained.

\hyphenation{met-al-li-ci-ty}
\subsubsection{Erroneous extinction, age and metallicity}
In (V,V--I) CMDs a strong correlation between 
extinction, age and metallicity exists (see also Ng{\muspc\&\muspc}Bertelli 
\cite{NB96} and references cited therein). 
A higher value of the extinction can be compensated
by a younger age and/or a lower metallicity.
Indeed, the results in Table~\ref{fixwrong_tbl} show that this
actually occurs for the lower age and metallicity limit.
The upper age and metallicity limit, however, drifts away in the 
opposite direction to compensate 
for `erroneous corrections' applied by other parameters.

A higher value of the star formation index results in 
a lower number of stars at the upper age metallicity limit.
To get a sufficient number of high metallicity stars    
one has to stretch the upper metallicity limit to a 
slightly higher value. 
\par
We further notice that a wrong value for the age and metallicity does not
affect the extinction significantly.
Our findings indirectly supports the method to determine 
high resolution ($4\arcmin\times4\arcmin$) 
extinction maps towards the Galactic bulge by Schultheis
et~al. (\cite{Schultheis99ea}) with the data obtained for the DeNIS project 
(Epchtein et~al.\ \cite{DeNIS97}). 

\subsubsection{Erroneous slope for power-law IMF}
The slope of the power-law IMF is 
very strongly constrained (Ng~\cite{Ng98a}) for the test population.
As a consequence the changes in the values of the 
remaining parameters are not extremely large.
The slightly larger value of the slope $\alpha$
pushes slightly more synthetic stars to fainter magnitudes,
introducing a relative deficiency of stars at brighter magnitudes.
This is compensated through a younger age, a decrease 
of the lower metallicity limit and an increase of the upper
metallicity limit. The higher value for the upper age 
and metallicity limit compensates
in its turn for the overestimation of the exponential
star formation index.\\
A lower value of $\alpha$ is partly compensated for by lowering the
upper age limit, lowering the lower metallicity limit  and
overestimating the upper metallicity limit.

\subsubsection{Erroneous index for exponential SFR}
A larger index for an exponentially decreasing SFR pushes more stars 
of the population to the blue edge of the CMD, 
resulting in a slightly bluer stellar population. \AMORE\  compensates 
this by mainly increasing the upper metallicity limit,
i.e. reddening the synthetic stellar population.\\
A lower value for the SFR index has the opposite effect. \AMORE\
compensates for the now slightly redder population by lowering 
the upper metallicity limit, making the population bluer.

\begin{table*}[t]
\caption{Description of the diagnostic statistics for simulations with
setup parameters from model \ref{first}-40. One of the parameters is
forced to a value  $\pm\muspc1\sigma$ from its original value. 
See Sects.~\ref{results_test4} and \ref{fitness} for additional details
\label{tbl3}}
\begin{center}
\begin{tabular}{|lc||ccc|ccc||c|}
\hline
model & offset & $N_{O,not}$ & $N_{S,not}$ & $N_{match}$ & $F_\chi$ & $F_P$ 
& $F$ & $f_A$ \\ 
\hline
\hline
ideal & & 0   &   0  & 5000 & 0.000 & 0.000 & 0.000 & 1.000 \cr
free  & & 60  &  60  & 4940 & 0.847 & 0.849 & 1.438 & 0.410 \cr
\hline
\hline
fixed: log d (pc) & $-1\sigma$   & 91 & 91 & 4909 & 0.788 & 1.287 & 2.278 & 0.305\cr
& $+1\sigma$     & 85 & 85 & 4915 & 0.722 & 1.202 & 1.967 & 0.337\cr
fixed: A$_V$ & $+1\sigma$          & 98 & 98 & 4902 & 0.932 & 1.386 &  3.790 & 0.264 \cr
fixed: $\log t_{low}$(yr) & $-1\sigma$ & 91 & 91 & 4901 & 0.897 & 1.287 & 2.460 &
0.289 \cr
 & $+1\sigma$   & 94 & 94 & 4906 & 0.905 & 1.329 & 2.587 &
0.278 \cr
fixed: $\log t_{hgh}$(yr) & $-1\sigma$ & 106  & 106 & 4894 & 0.977 & 1.499 & 3.202 &
0.238 \cr
& $+1\sigma$   & 126  & 126 & 4874 & 1.017 & 1.782 & 4.211 &
0.192 \cr
fixed: $[{\rm Z}]_{low}$  & $-1\sigma$ & 78 & 78 & 4922 & 0.907 & 1.103 & 2.043 &
0.329 \cr
& $+1\sigma$   & 80 & 80 & 4920 & 0.835 & 1.131 & 1.977 &
0.336 \cr
fixed: $[{\rm Z}]_{hgh}$  & $-1\sigma$ & 148 & 148 & 4852 & 0.896 & 2.094 & 5.188 &
0.162 \cr
& $+1\sigma$   & 94 & 94 & 4906 & 0.867 & 1.329 & 2.519 &
0.284 \cr
fixed: $\alpha$ & $-1\sigma$ &  95 &  95 & 4905 & 0.886 & 1.344 & 2.590 & 0.279 \cr
& $+1\sigma$   & 103 & 103 & 4897 & 0.894 & 1.457 & 2.921 & 0.255 \cr
fixed: $\beta$  & $-1\sigma$ & 145 & 145 & 4855 & 0.930 & 2.051 & 5.071 & 0.165 \cr
& $+1\sigma$   & 103 & 103 & 4897 & 0.834 & 1.457 & 2.818 & 0.262 \cr
\hline
\end{tabular}
\end{center}
\end{table*}

\section{Discussion}
\label{Discussion}

\subsection{Relative contribution of the parameters to the fitness}
Figure~\ref{evolveamorefig} and the Tables~\ref{tbl1}, 
\ref{statgood_tbl}, \ref{statwrong_tbl}, \ref{fixgood_tbl} 
\& \ref{fixwrong_tbl}
provide the hint that not all parameters have an equal contribution to the
fitness. 
It appears that the $\beta$ and Z$_{high}$ parameter can vary considerably
and still yield a decent fitness. Moreover, 
Tables~\ref{fixgood_tbl} \& \ref{statgood_tbl} indicate that
knowledge of the value of the Z$_{high}$ parameter results in acceptable values
for the other parameters.\\
The origin of this behaviour lies in the implicit definition of the 
exponential star formation rate (for $\beta\!=\!1$ one has a decreasing star 
formation towards a younger age) attached to a linear age-metallicity
relation. The latter relation will give less metal-richer stars. 
The small number of stars with higher metallicity induces a larger
variation of the Z$_{high}$ parameter without affecting significantly 
the overall fitness.

\subsection{Convergence}
\label{convergence}
The fine-tuning of the genetic algorithm is a tedious task. 
It is not straightforward to find the optimum setting for the
problem to be solved. One has to balance the exploring quality
through crossovers
against the variation of the parameters through (creep) mutations. 
\hfill\break
We did not want to deal with a mutation dominated search,
because it tends to move farther away from an optimum parameter setting
in the majority of the cases. 
We used therefore a relatively high crossover probability ({\tt pcross})
and we set the mutations at a fixed rate, such that 
on average only 2.8 mutations occur in the gene pool of each individual.
\par
\hyphenation{im-pro-ve-ment}
At a certain stage however one requires the variation
of other correlated parameters
to obtain an improvement.  
This becomes particularly necessary when approaching the 
optimum setting of the parameters. 
A favourable crossover and mutation might do the trick, but 
it might take a while before this occurs. 
We introduced in Sect.~\ref{correlated}
the possibility that two parameters might
be more sensitive to mutations than others.
This approach gave better results for the majority of the trial cases
(see Table~\ref{avgfit}), 
but it failed to obtain improvements  
when changes of one parameter were neutralized through 
the variation of one or more parameters.
The distance-extinction and the age-metallicity degeneracies slow down
the convergence of \AMORE\ for $f\!>\!0.3$, see Fig.~\ref{evolveamorefig}.
\par
One of the modifications to consider for future implementation
is a two-chromosome approach. In that case
acceptable values for the parameters do not shift out of the
population if the overall fitness is less, but still 
reside in the gene pool as a recessive quality. This however, will
require a major extension to \PIKAIA\ and a significant amount
of genetic research to be done about dominant and recessive qualities
in the \AMORE\ gene pool.\\
Another modification to consider in order to improve the accuracy and
to speed up convergence, is to replace the finite resolution of the
digital encoding scheme with a genetic coding based on floating point,
i.e. each gene on the chromosome is represented by one floating point
number. According to Michalewicz (\cite{Michalewicz96}) a real encoding
scheme can be superior and improve convergence.
Such an encoding scheme is indeed to be included in the
next release of \PIKAIA\ 2.0 (Charbonneau; in preparation).

\subsubsection{Unstable solutions}
\label{unstable}
In one test (fixing A$_V$ at one sigma above the original value for
model 9) no convergence was achieved and the run was aborted. 
Because \AMORE\ is quite sensitive to rounding these effects can be
circumvented by slightly altering the input parameters. We decided
in this case against such an action, because that would make the sample
inhomogeneous.

\subsection{Degeneracy}
\label{degeneracy}
Isochrones for a particular age and metallicity can be mimicked with
another set of isochrones of different age and metallicity
(Worthey \cite{Worthey94}, Charlot et~al.\ \cite{Charlot96ea},
and references cited therein). This degeneracy of the parameter space
increases if one considers the distance and the extinction
towards a stellar aggregate. There is no straight forward method
to circumvent partial degeneracy of the parameters to be explored.
One might consider to apply \AMORE\  for the analysis of 
colour-colour diagrams in order to
rule out the distance, to determine the extinction \& 
a number of other parameters,
and finally to determine the distance to the stellar population
from one of the CMDs.
The combined analysis of colour-magnitude and colour-colour diagrams
is expected to improve the results obtained by \AMORE\ so far.
However, this requires that at least one additional colour should be 
available for each star considered above a certain detection
and completeness threshold. Moreover, as was mentioned in
Sect.~\ref{res_fixageAv}, knowledge of the extinction does not
automatically imply that the distance can be retrieved accurately.
\par
As demonstrated in Sect.~\ref{res_test1} the degeneracy among parameters 
becomes noticeable for $f\!>\!0.25$ or $F\!<\!3$.
This corresponds to 
{systematic offset }
for each parameter
of on average $\sim0.6\sigma_k$ and at maximum $\sim1\sigma_k$.
The Poisson uncertainty of the original population results in
a fitness of $f\!=\!0.43$ ($F\!\simeq\!1.33$).
However, solutions with a comparable fitness do exist due
to the degeneracy of the parameter space. 
A direct consequence is that there is an intrinsic 
{offset }
present among the parameters amounting to
on average $\sim0.4\sigma_k$ and at maximum $\sim0.7\sigma_k$.
This intrinsic 
{offset }
is present in the solutions obtained 
with \AMORE\  and actually is responsible for slowing down the convergence 
in the fitness range $0.30\!<\!f\!<\!0.43$.
It will therefore be nearly impossible to
recover in one pass the original input values. 
However, some improvements might be obtained by averaging the parameter
values obtained from \AMORE\ runs with different initial conditions.

\section{Conclusions}
\label{conclusions}
We demonstrate that an automatic search can be made 
for the astrophysical parameters of a synthetic stellar population 
from the analysis of colour-magnitude diagrams with an optimizer, 
based on a genetic algorithm.
However, \AMORE\ tends to slightly underestimate
the distance. It subsequently attempts to compensate this with
an higher extinction, a higher upper age and a slightly
steeper slope for the power-law IMF.
At $f\!>\!0.3$ the combined effect of the age-metallicity \& the
distance-extinction degeneracy slows down the convergence. The data
suggests that \AMORE\ has more problems dealing with the
age-metallicity than with the distance-extinction degeneracy.
For general purpose, however, the retrieved values
are sufficiently accurate.

\section{Future work}
\label{future}
The good results obtained so far for a single synthetic stellar
population is an indication about \AMORE's potential for
the detailed analysis of CMDs. { The next step is to improve one step at a time
various aspects of \AMORE\ before it can be used as an
interpretative tool for large photometric surveys.}
\\
Despite limitations
in the input physics of the underlying stellar evolutionary tracks and 
by the transformation from the theoretical to the 
observational plane, the results with real data 
from Gallart et~al.\ (\cite{Gallart99ea}),
who uses the same set of evolutionary tracks, are encouraging.
It will therefore be important to 
verify first with, for example, well studied open clusters 
(see Carraro et~al.\ \cite{CNP98, Carraro99ea} 
and references cited therein) 
for which age and metallicity range we may apply \AMORE\  safely. 
Extinction is also of some concern, because a high extinction may result
in the MS turnoff point to fall below the detection limit. This would
deprive \AMORE\ of a clear reference point.
\par
Another case of interest is of course the question how many
different stellar populations can be distinguished with \AMORE.
Separating multiple, mixed populations from each other through
the automated and the objective analysis of colour-magnitude
diagrams could be a valuable tool for the analysis of galaxy formation
and evolution.\hfill\break
This requires a rigorous follow up study on the
separation of multiple (synthetic) populations.
One further ought to verify if the automated analysis
of colour-colour diagram can reduce the effects of error
cancellation between distance and extinction.
\par
{
Finally, after a succesful implementation, testing and validation phase,
we plan to combine \AMORE\  with the Padova 
spectrophotometric code (see Bressan et~al.\ \cite{Bres94ea}, \cite{Bres96ea};
Tantalo et~al.\ \cite{Tant96ea}, 1998ab). 
A synthetic population has to be generated, containing sub-populations
with different ages and metallicities. Then a synthetic spectrum must
be generated for the mixed population and subsequently used as input
for a synthetic, spectral fitting program to determine the underlying stellar
populations. In this way one can establish the calibration of the
spectrophotometric tool in a self-consistent way. Furthermore, an
implicit verification can be made that the populations are consistent
with those obtained from a CMD analysis with \AMORE.}

\begin{acknowledgements}{
The research was partly supported by TMR grant 
ERBFMRX-CT96-0086 from the European Community
(Network: Formation and Evolution of Galaxies), by the Italian
Ministry of University, Scientific Research and Technology (MURST) and by the
Italian Space Agency (ASI).
}
\end{acknowledgements}

\appendix
\section{\AMORE\ test results, the data}
The appendix contains the data of the simulations as discussed 
in Sect.~\ref{test}. The first test, in which we explored different
values for the \PIKAIA\ control parameters, is described in
Sect.~\ref{desc_test1} and the data are given in Table~\ref{first}.\\
The data of the third test, in which one of the astrophysical parameters was fixed
at its correct value as discussed in Sect.~\ref{desc_test3}, are given
in Table~\ref{fixgood_tbl}.
The data on the fourth test, in which one parameter was fixed one
sigma from its correct value as discussed in Sect.~\ref{desc_test4},
are given in Table~\ref{fixwrong_tbl}.
\\
The tables in the Appendix are available in electronic form at the CDS.

\newpage
\onecolumn
\begin{longtable}{|r|ccc|cc|cc|}
\caption{\small
\hsize=16.50cm\AMORE\ with the test population as specified in
Sect.~\ref{desc_test1}. See \ref{crossover}, \ref{brood}, \ref{creep},
\ref{correlated} and \ref{setup} for additional
details. {\tt pcross} is the crossover probability, {\tt rcross} is
the multi-point crossover rate, {\tt rbrood} determines the amount of
offspring two parents can have, {\tt pcreep} is the creep mutation
rate and {\tt pcorr} the correlated mutation rate. The generation
number indicates at which generation the resulting fitness value $f_A$
obtained with \AMORE\ first emerged}
\label{first}\\
\hline
model & {\tt pcross} & {\tt rcross} & {\tt rbrood } & {\tt pcreep} &
{\tt pcorr} & $f_A$ & generation\\
\hline
\endfirsthead
\caption{continued} \\
\hline
model & {\tt pcross} & {\tt rcross} & {\tt rbrood }  
& {\tt pcreep} & {\tt pcorr} & $f_A$ & generation\\ 
\hline
\endhead
\hline
\endfoot
1   & 0.85 & 1.00 & 1.00 & 0.0 & 0.0 & 0.17266  & 361\\
2   & 0.85 & 1.00 & 1.00 & 0.3 & 0.0 & 0.17266	& 361\\
3   & 0.85 & 1.00 & 1.00 & 0.7 & 0.0 & 0.17266	& 361\\
4   & 0.85 & 1.00 & 1.00 & 0.0 & 0.3 & 0.27551	& 221\\
5   & 0.85 & 1.00 & 1.00 & 0.0 & 0.7 & 0.26830	& 221\\
6   & 0.85 & 1.00 & 1.00 & 0.3 & 0.3 & 0.30488	& 381\\
7   & 0.85 & 1.00 & 1.00 & 0.7 & 0.3 & 0.32005	& 341\\
8   & 0.85 & 1.00 & 1.00 & 0.3 & 0.7 & 0.32167	& 201\\
9   & 0.85 & 1.00 & 1.00 & 0.7 & 0.7 & 0.38310	& 341\\
10  & 0.50 & 1.00 & 1.00 & 0.0 & 0.0 & 0.35662	& 381\\
11  & 0.50 & 1.00 & 1.00 & 0.3 & 0.0 & 0.35662	& 381\\
12  & 0.50 & 1.00 & 1.00 & 0.7 & 0.0 & 0.35662	& 381\\
13  & 0.50 & 1.00 & 1.00 & 0.0 & 0.3 & 0.34961	& 281\\
14  & 0.50 & 1.00 & 1.00 & 0.0 & 0.7 & 0.17008	& 261\\
15  & 0.50 & 1.00 & 1.00 & 0.3 & 0.3 & 0.36025	& 381\\
16  & 0.50 & 1.00 & 1.00 & 0.7 & 0.3 & 0.24048	& 261\\
17  & 0.50 & 1.00 & 1.00 & 0.3 & 0.7 & 0.23956	& 321\\
18  & 0.50 & 1.00 & 1.00 & 0.7 & 0.7 & 0.29463	& 321\\
19  & 0.85 & 2.00 & 1.00 & 0.0 & 0.0 & 0.31700	& 384\\
20  & 0.85 & 2.00 & 1.00 & 0.3 & 0.0 & 0.31700	& 384\\
21  & 0.85 & 2.00 & 1.00 & 0.7 & 0.0 & 0.31700	& 384\\
22  & 0.85 & 2.00 & 1.00 & 0.0 & 0.3 & 0.25740	& 341\\
23  & 0.85 & 2.00 & 1.00 & 0.0 & 0.7 & 0.35805	& 341\\
24  & 0.85 & 2.00 & 1.00 & 0.3 & 0.3 & 0.34120	& 181\\
25  & 0.85 & 2.00 & 1.00 & 0.7 & 0.3 & 0.31290	& 141\\
26  & 0.85 & 2.00 & 1.00 & 0.3 & 0.7 & 0.30913	& 341\\
27  & 0.85 & 2.00 & 1.00 & 0.7 & 0.7 & 0.32212	& 261\\
28  & 0.50 & 2.00 & 1.00 & 0.0 & 0.0 & 0.16122	& 261\\
29  & 0.50 & 2.00 & 1.00 & 0.3 & 0.0 & 0.16122	& 261\\
30  & 0.50 & 2.00 & 1.00 & 0.7 & 0.0 & 0.16122	& 261\\
31  & 0.50 & 2.00 & 1.00 & 0.0 & 0.3 & 0.36796	& 381\\
32  & 0.50 & 2.00 & 1.00 & 0.0 & 0.7 & 0.28494	& 381\\
33  & 0.50 & 2.00 & 1.00 & 0.3 & 0.3 & 0.27263	& 381\\
34  & 0.50 & 2.00 & 1.00 & 0.7 & 0.3 & 0.16352	& 301\\
35  & 0.50 & 2.00 & 1.00 & 0.3 & 0.7 & 0.35149	& 301\\
36  & 0.50 & 2.00 & 1.00 & 0.7 & 0.7 & 0.34031	& 381\\
37  & 0.85 & 3.00 & 1.00 & 0.0 & 0.0 & 0.38297	& 301\\
38  & 0.85 & 3.00 & 1.00 & 0.3 & 0.0 & 0.38297	& 301\\
39  & 0.85 & 3.00 & 1.00 & 0.7 & 0.0 & 0.38297	& 301\\
40  & 0.85 & 3.00 & 1.00 & 0.0 & 0.3 & 0.41019	& 341\\
41  & 0.85 & 3.00 & 1.00 & 0.0 & 0.7 & 0.32012	& 381\\
42  & 0.85 & 3.00 & 1.00 & 0.3 & 0.3 & 0.36840	& 341\\
43  & 0.85 & 3.00 & 1.00 & 0.7 & 0.3 & 0.31281	& 361\\
44  & 0.85 & 3.00 & 1.00 & 0.7 & 0.3 & 0.31281	& 361\\
45  & 0.85 & 3.00 & 1.00 & 0.3 & 0.7 & 0.30571	& 381\\
46  & 0.50 & 3.00 & 1.00 & 0.0 & 0.0 & 0.15614	& 301\\
47  & 0.50 & 3.00 & 1.00 & 0.3 & 0.0 & 0.15614	& 301\\
48  & 0.50 & 3.00 & 1.00 & 0.7 & 0.0 & 0.15614	& 301\\
49  & 0.50 & 3.00 & 1.00 & 0.0 & 0.3 & 0.23392	& 301\\
50  & 0.50 & 3.00 & 1.00 & 0.0 & 0.7 & 0.16274	& 221\\
51  & 0.50 & 3.00 & 1.00 & 0.3 & 0.3 & 0.25123	& 281\\
52  & 0.50 & 3.00 & 1.00 & 0.7 & 0.3 & 0.27151	& 261\\
53  & 0.50 & 3.00 & 1.00 & 0.3 & 0.7 & 0.35597	& 301\\
54  & 0.50 & 3.00 & 1.00 & 0.7 & 0.7 & 0.27743	& 361\\
55  & 0.85 & 1.00 & 2.00 & 0.0 & 0.0 & 0.36221	& 161\\
56  & 0.85 & 1.00 & 2.00 & 0.3 & 0.0 & 0.36221	& 161\\
57  & 0.85 & 1.00 & 2.00 & 0.7 & 0.0 & 0.36221	& 161\\
58  & 0.85 & 1.00 & 2.00 & 0.0 & 0.3 & 0.16284	& 261\\
59  & 0.85 & 1.00 & 2.00 & 0.0 & 0.7 & 0.30251	& 381\\
60  & 0.85 & 1.00 & 2.00 & 0.3 & 0.3 & 0.32133	& 361\\
61  & 0.85 & 1.00 & 2.00 & 0.7 & 0.3 & 0.33565	& 361\\
62  & 0.85 & 1.00 & 2.00 & 0.3 & 0.7 & 0.27452	& 341\\
63  & 0.85 & 1.00 & 2.00 & 0.7 & 0.7 & 0.34660	& 301\\
64  & 0.50 & 1.00 & 2.00 & 0.0 & 0.0 & 0.27683	& 381\\
65  & 0.50 & 1.00 & 2.00 & 0.3 & 0.0 & 0.27683	& 381\\
66  & 0.50 & 1.00 & 2.00 & 0.7 & 0.0 & 0.27683	& 381\\
67  & 0.50 & 1.00 & 2.00 & 0.0 & 0.3 & 0.16833	& 201\\
68  & 0.50 & 1.00 & 2.00 & 0.0 & 0.7 & 0.38586	& 321\\
69  & 0.50 & 1.00 & 2.00 & 0.3 & 0.3 & 0.34390	& 361\\
70  & 0.50 & 1.00 & 2.00 & 0.7 & 0.3 & 0.33531	& 241\\
71  & 0.50 & 1.00 & 2.00 & 0.3 & 0.7 & 0.32646	& 221\\
72  & 0.50 & 1.00 & 2.00 & 0.7 & 0.7 & 0.34591	& 381\\
73  & 0.85 & 1.00 & 4.00 & 0.0 & 0.0 & 0.34607	& 221\\
74  & 0.85 & 1.00 & 4.00 & 0.3 & 0.0 & 0.34607	& 221\\
75  & 0.85 & 1.00 & 4.00 & 0.7 & 0.0 & 0.34607	& 221\\
76  & 0.85 & 1.00 & 4.00 & 0.0 & 0.3 & 0.32143	& 361\\
77  & 0.85 & 1.00 & 4.00 & 0.0 & 0.7 & 0.27159	& 381\\
78  & 0.85 & 1.00 & 4.00 & 0.3 & 0.3 & 0.26001	& 221\\
79  & 0.85 & 1.00 & 4.00 & 0.7 & 0.3 & 0.19802	& 241\\
80  & 0.85 & 1.00 & 4.00 & 0.3 & 0.7 & 0.33032	& 381\\
81  & 0.85 & 1.00 & 4.00 & 0.7 & 0.7 & 0.17799	& 341\\
82  & 0.50 & 1.00 & 4.00 & 0.0 & 0.0 & 0.27728	& 201\\
83  & 0.50 & 1.00 & 4.00 & 0.3 & 0.0 & 0.27728	& 201\\
84  & 0.50 & 1.00 & 4.00 & 0.7 & 0.0 & 0.27728	& 201\\
85  & 0.50 & 1.00 & 4.00 & 0.0 & 0.3 & 0.35783	& 381\\
86  & 0.50 & 1.00 & 4.00 & 0.0 & 0.7 & 0.28070	& 341\\
87  & 0.50 & 1.00 & 4.00 & 0.3 & 0.3 & 0.33717	& 381\\
88  & 0.50 & 1.00 & 4.00 & 0.7 & 0.3 & 0.18275	& 341\\
89  & 0.50 & 1.00 & 4.00 & 0.3 & 0.7 & 0.24795	& 361\\
90  & 0.50 & 1.00 & 4.00 & 0.7 & 0.7 & 0.28384	& 341\\
91  & 0.85 & 2.00 & 2.00 & 0.0 & 0.0 & 0.29274	& 261\\
92  & 0.85 & 2.00 & 2.00 & 0.3 & 0.0 & 0.29274	& 261\\
93  & 0.85 & 2.00 & 2.00 & 0.7 & 0.0 & 0.29274	& 261\\
94  & 0.85 & 2.00 & 2.00 & 0.0 & 0.3 & 0.24259	& 381\\
95  & 0.85 & 2.00 & 2.00 & 0.0 & 0.7 & 0.26513	& 161\\
96  & 0.85 & 2.00 & 2.00 & 0.3 & 0.3 & 0.33411	& 361\\
97  & 0.85 & 2.00 & 2.00 & 0.7 & 0.3 & 0.25808	& 121\\
98  & 0.85 & 2.00 & 2.00 & 0.3 & 0.7 & 0.35686	& 341\\
99  & 0.85 & 2.00 & 2.00 & 0.7 & 0.7 & 0.37254	& 399\\
100 & 0.50 & 2.00 & 2.00 & 0.0 & 0.0 & 0.28259	& 361\\
101 & 0.50 & 2.00 & 2.00 & 0.3 & 0.0 & 0.28259	& 361\\
102 & 0.50 & 2.00 & 2.00 & 0.7 & 0.0 & 0.28259	& 361\\
103 & 0.50 & 2.00 & 2.00 & 0.0 & 0.3 & 0.34559	& 341\\
104 & 0.50 & 2.00 & 2.00 & 0.0 & 0.7 & 0.39464	& 320\\
105 & 0.50 & 2.00 & 2.00 & 0.3 & 0.3 & 0.27112	& 181\\
106 & 0.50 & 2.00 & 2.00 & 0.7 & 0.3 & 0.29866	& 321\\
107 & 0.50 & 2.00 & 2.00 & 0.3 & 0.7 & 0.23660	& 181\\
108 & 0.50 & 2.00 & 2.00 & 0.7 & 0.7 & 0.35769	& 261\\
109 & 0.85 & 2.00 & 4.00 & 0.0 & 0.0 & 0.32628	& 181\\
110 & 0.85 & 2.00 & 4.00 & 0.3 & 0.0 & 0.32628	& 181\\
111 & 0.85 & 2.00 & 4.00 & 0.7 & 0.0 & 0.32628	& 181\\
112 & 0.85 & 2.00 & 4.00 & 0.0 & 0.3 & 0.26910	& 361\\
113 & 0.85 & 2.00 & 4.00 & 0.0 & 0.7 & 0.37278	& 389\\
114 & 0.85 & 2.00 & 4.00 & 0.3 & 0.3 & 0.31434	& 241\\
115 & 0.85 & 2.00 & 4.00 & 0.7 & 0.3 & 0.18222	& 321\\
116 & 0.85 & 2.00 & 4.00 & 0.3 & 0.7 & 0.26176	& 341\\
117 & 0.85 & 2.00 & 4.00 & 0.7 & 0.7 & 0.19495	& 241\\
118 & 0.50 & 2.00 & 4.00 & 0.0 & 0.0 & 0.17656	& 361\\
119 & 0.50 & 2.00 & 4.00 & 0.3 & 0.0 & 0.17656	& 361\\
120 & 0.50 & 2.00 & 4.00 & 0.7 & 0.0 & 0.17656	& 361\\
121 & 0.50 & 2.00 & 4.00 & 0.0 & 0.3 & 0.17248	& 181\\
122 & 0.50 & 2.00 & 4.00 & 0.0 & 0.7 & 0.27790	& 381\\
123 & 0.50 & 2.00 & 4.00 & 0.3 & 0.3 & 0.17117	& 221\\
124 & 0.50 & 2.00 & 4.00 & 0.7 & 0.3 & 0.36171	& 381\\
125 & 0.50 & 2.00 & 4.00 & 0.3 & 0.7 & 0.38431	& 392\\
126 & 0.50 & 2.00 & 4.00 & 0.7 & 0.7 & 0.31711	& 400\\
127 & 0.85 & 3.00 & 2.00 & 0.0 & 0.0 & 0.16288	& 161\\
128 & 0.85 & 3.00 & 2.00 & 0.3 & 0.0 & 0.16288	& 161\\
129 & 0.85 & 3.00 & 2.00 & 0.7 & 0.0 & 0.16288	& 161\\
130 & 0.85 & 3.00 & 2.00 & 0.0 & 0.3 & 0.34149	& 341\\
131 & 0.85 & 3.00 & 2.00 & 0.0 & 0.7 & 0.33630	& 281\\
132 & 0.85 & 3.00 & 2.00 & 0.3 & 0.3 & 0.24355	& 301\\
133 & 0.85 & 3.00 & 2.00 & 0.7 & 0.3 & 0.33731	& 221\\
134 & 0.85 & 3.00 & 2.00 & 0.3 & 0.7 & 0.26283	& 201\\
135 & 0.85 & 3.00 & 2.00 & 0.7 & 0.7 & 0.25975	& 121\\
136 & 0.50 & 3.00 & 2.00 & 0.0 & 0.0 & 0.26047	& 221\\
137 & 0.50 & 3.00 & 2.00 & 0.3 & 0.0 & 0.26047	& 221\\
138 & 0.50 & 3.00 & 2.00 & 0.7 & 0.0 & 0.26047	& 221\\
139 & 0.50 & 3.00 & 2.00 & 0.0 & 0.3 & 0.26449	& 321\\
140 & 0.50 & 3.00 & 2.00 & 0.0 & 0.7 & 0.27116	& 281\\
141 & 0.50 & 3.00 & 2.00 & 0.3 & 0.3 & 0.31298	& 101\\
142 & 0.50 & 3.00 & 2.00 & 0.7 & 0.3 & 0.32952	& 201\\
143 & 0.50 & 3.00 & 2.00 & 0.3 & 0.7 & 0.25496	& 281\\
144 & 0.50 & 3.00 & 2.00 & 0.7 & 0.7 & 0.24888	& 221\\
145 & 0.85 & 3.00 & 4.00 & 0.0 & 0.0 & 0.34932	& 321\\
146 & 0.85 & 3.00 & 4.00 & 0.3 & 0.0 & 0.34932	& 321\\
147 & 0.85 & 3.00 & 4.00 & 0.7 & 0.0 & 0.34932	& 321\\
148 & 0.85 & 3.00 & 4.00 & 0.0 & 0.3 & 0.34687	& 381\\
149 & 0.85 & 3.00 & 4.00 & 0.0 & 0.7 & 0.32433	& 321\\
150 & 0.85 & 3.00 & 4.00 & 0.3 & 0.3 & 0.17152	& 141\\
151 & 0.85 & 3.00 & 4.00 & 0.7 & 0.3 & 0.30958	& 395\\
152 & 0.85 & 3.00 & 4.00 & 0.3 & 0.7 & 0.31341	& 398\\
153 & 0.85 & 3.00 & 4.00 & 0.7 & 0.7 & 0.27072	& 261\\
154 & 0.50 & 3.00 & 4.00 & 0.0 & 0.0 & 0.34272	& 101\\
155 & 0.50 & 3.00 & 4.00 & 0.3 & 0.0 & 0.34272	& 101\\
156 & 0.50 & 3.00 & 4.00 & 0.7 & 0.0 & 0.34272	& 101\\
157 & 0.50 & 3.00 & 4.00 & 0.0 & 0.3 & 0.30844	& 321\\
158 & 0.50 & 3.00 & 4.00 & 0.0 & 0.7 & 0.27009	& 361\\
159 & 0.50 & 3.00 & 4.00 & 0.3 & 0.3 & 0.17333	& 381\\
160 & 0.50 & 3.00 & 4.00 & 0.7 & 0.3 & 0.25621	& 361\\
161 & 0.50 & 3.00 & 4.00 & 0.3 & 0.7 & 0.39429	& 341\\
162 & 0.50 & 3.00 & 4.00 & 0.7 & 0.7 & 0.26150	& 361\\
\hline
\end{longtable}
\begin{table}[h!]
\caption{Results of running \AMORE\ with one of 
the astrophysical
parameters fixed at its original value. These tests provide an
indication about the influence of one parameter on the retrieval
of the remaining parameters,
see Sect.~\ref{desc_test3} and \ref{results_test3} for
additional details}
\label{fixgood_tbl}
\begin{center}
\begin{tabular}{|l|r|c|c|c|c|c|c|c|c|c|c|}
\hline
parameter & model & $f_A$ & log d(pc)& A$_V$ & 
$\log t_{low}$ & $\log t_{hgh}$ & [Z]$_{low}$ & [Z]$_{hgh}$ &
$\alpha$ & $\beta$\\
\hline
\hline
log d          & 9  & 0.34897   & 3.90633  &  0.002  &  9.89178  &  9.95342  &  \minus0.60089  &  0.19300  &  2.341  & 0.985\\
	       & 14 & 0.35919   & 3.90633  & \minus0.001  &  9.88937  &  9.95245  &  \minus0.60115  &  0.20684  &  2.341  & 1.111\\
	       & 22 & 0.26617   & 3.90633  &  0.007  &  9.86685  &  9.95757  &  \minus0.53389  &  0.30054  &  2.351  & 2.980\\
	       & 34 & 0.31373   & 3.90633  &  0.001  &  9.87873  &  9.95335  &  \minus0.57392  &  0.25198  &  2.341  & 1.613\\
	       & 40 & 0.15102   & 3.90633  &  0.016  &  9.78469  & 10.12246  &  \minus0.69199  &  0.07931  &  2.319  & \minus1.281\\
	       & 52 & 0.35475   & 3.90633  &  0.012  &  9.89392  &  9.95002  &  \minus0.59523  &  0.17305  &  2.338  & 0.936\\
\hline
A$_V$          & 9  & 0.25192   & 3.89083  &  0.000  &  9.87519  &  9.98656  &  \minus0.50206  &  0.34423  &  2.391  & 2.613\\
	       & 14 & 0.23362   & 3.88471  &  0.000  &  9.88175  &  9.99802  &  \minus0.50635  &  0.40461  &  2.424  & 3.682\\
	       & 22 & 0.30930   & 3.89399  & \minus0.001  &  9.91339  &  9.98285  &  \minus0.55292  &  0.20319  &  2.383  & 0.889\\
	       & 34 & 0.17545   & 3.89688  &  0.001  &  9.80693  &  10.1395  &  \minus0.66073  &  0.10056  &  2.339  & \minus1.338\\
	       & 40 & 0.26131   & 3.89634  &  0.001  &  9.82335  &  9.96459  &  \minus0.51903  &  0.44994  &  2.355  & 3.493\\
	       & 52 & 0.41406   & 3.90592  &  0.000  &  9.90287  &  9.95544  &  \minus0.59621  &  0.17710  &  2.350  & 1.003\\
\hline
log t$_{low}$  &  9 & 0.34505   & 3.89687  &  0.030  &  9.90310  & 9.96292  & \minus0.59785  & 0.19586  & 2.358  & 1.028\\
	       & 14 & 0.29609   & 3.90078  &  0.064  &  9.90308  & 9.93943  & \minus0.67152  & 0.10196  & 2.333  & 0.087\\
	       & 22 & 0.34452   & 3.89289  &  0.026  &  9.90308  & 9.96482  & \minus0.57260  & 0.23664  & 2.362  & 1.435\\
	       & 34 & 0.32855   & 3.89861  &  0.067  &  9.90309  & 9.94303  & \minus0.62670  & 0.14375  & 2.343  & 0.845\\
	       & 40 & 0.35429   & 3.89657  &  0.014  &  9.90309  & 9.96685  & \minus0.56333  & 0.21007  & 2.358  & 1.130\\
	       & 52 & 0.32296   & 3.89388  &  0.036  &  9.90309  & 9.96827  & \minus0.58340  & 0.21130  & 2.365  & 1.197\\
\hline
log t$_{high}$ & 9  & 0.32414   & 3.89839  &  0.033  &  9.88329  & 9.95424  & \minus0.55256  & 0.26193  & 2.341  & 1.674\\
	       & 14 & 0.25502   & 3.89455  &  0.036  &  9.85431  & 9.95425  & \minus0.52027  & 0.39343  & 2.351  & 3.455\\
	       & 22 & 0.26055   & 3.90038  &  0.022  &  9.76480  & 9.95423  & \minus0.52358  & 0.52141  & 2.333  & 4.267\\
	       & 34 & 0.36575   & 3.89772  &  0.046  &  9.90052  & 9.95424  & \minus0.60081  & 0.20274  & 2.352  & 1.251\\
	       & 40 & 0.25141   & 3.89839  &  0.042  &  9.75973  & 9.95424  & \minus0.56527  & 0.52340  & 2.335  & 4.878\\
	       & 52 & 0.37074   & 3.89458  &  0.066  &  9.91101  & 9.95424  & \minus0.62634  & 0.16462  & 2.361  & 1.046\\
\hline
log Z$_{low}$  & 9  & 0.22556   & 3.89237  &  0.059  &  9.78605  & 9.96444  & \minus0.60205  & 0.48867  & 2.363  & 4.119\\
	       & 14 & 0.32714   & 3.89352  &  0.031  &  9.89643  & 9.96409  & \minus0.60207  & 0.22149  & 2.354  & 0.907\\
	       & 22 & 0.23902   & 3.89510  &  0.047  &  9.82960  & 9.95708  & \minus0.60205  & 0.45203  & 2.363  & 3.709\\
	       & 34 & 0.27029   & 3.89729  &  0.040  &  9.84880  & 9.95803  & \minus0.60207  & 0.31992  & 2.327  & 1.756\\
	       & 40 & 0.28851   & 3.89175  &  0.028  &  9.94114  & 9.97064  & \minus0.60206  & 0.10911  & 2.399  & 0.305\\
	       & 52 & 0.26238   & 3.89419  &  0.041  &  9.87970  & 9.96558  & \minus0.60207  & 0.29439  & 2.375  & 2.197\\
\hline
log Z$_{high}$ & 9  & 0.32971   & 3.89633  &  0.044  &  9.90542  & 9.96035  & \minus0.58239  & 0.17609  & 2.363  & 1.182\\
	       & 14 & 0.39924   & 3.89593  &  0.038  &  9.90880  & 9.96147  & \minus0.58144  & 0.17608  & 2.360  & 1.009\\
	       & 22 & 0.40156   & 3.89579  &  0.053  &  9.90555  & 9.95702  & \minus0.60422  & 0.17608  & 2.357  & 1.057\\
	       & 34 & 0.31199   & 3.89306  &  0.017  &  9.92288  & 9.97460  & \minus0.57681  & 0.17610  & 2.384  & 0.872\\
	       & 40 & 0.35466   & 3.89952  &  0.044  &  9.90467  & 9.95065  & \minus0.61730  & 0.17609  & 2.349  & 1.010\\
	       & 52 & 0.36582   & 3.89853  &  0.042  &  9.90077  & 9.95166  & \minus0.60430  & 0.17608  & 2.342  & 0.878\\
\hline
$\alpha$       & 9  & 0.34047	& 3.89758  &  0.023  &  9.89805  &  9.96360  &  \minus0.57367  &  0.19013  &  2.350  &  0.831\\
	       & 14 & 0.17505	& 3.89826  &  0.014  &  9.82400	 & 10.14663  &  \minus0.62474  &  0.04615  &  2.349  & \minus1.950\\
	       & 22 & 0.30768   & 3.89446  &  0.063  &  9.90882  &  9.95914  &  \minus0.73991  &  0.12715  &  2.349  & \minus0.035\\
	       & 34 & 0.31327   & 3.89614  &  0.019  &  9.87877  &  9.96638  &  \minus0.52373  &  0.26868  &  2.350  & 1.784\\
	       & 40 & 0.35728   & 3.89731  &  0.021  &  9.89219  &  9.96144  &  \minus0.55927  &  0.23623  &  2.349  & 1.347\\
	       & 52 & 0.33530   & 3.89702  &  0.027  &  9.89860  &  9.96481  &  \minus0.54426  &  0.19782  &  2.350  & 1.039\\
\hline
$\beta$        & 9  & 0.31038   & 3.89352  &  0.082  &  9.91755  &  9.94928  &  \minus0.62624  &  0.13301  &  2.360  & 1.000\\
	       & 14 & 0.38179   & 3.89881  &  0.033  &  9.90224  &  9.95851  &  \minus0.57609  &  0.17513  &  2.352  & 1.000\\
	       & 22 & 0.34095   & 3.89848  &  0.052  &  9.89348  &  9.95316  &  \minus0.57848  &  0.18922  &  2.338  & 0.999\\
	       & 34 & 0.16612   & 3.89537  &  0.028  &  9.80857  &  10.0490  &  \minus0.44888  &  0.12520  &  2.382  & 1.001\\
	       & 40 & 0.35606   & 3.90903  &  0.019  &  9.88940  &  9.94183  &  \minus0.61022  &  0.17903  &  2.329  & 1.001\\
	       & 52 & 0.31834   & 3.89857  &  0.073  &  9.91742  &  9.93856  &  \minus0.61263  &  0.12154  &  2.350  & 1.000\\
\hline
\end{tabular}
\end{center}
\end{table}
{\scriptsize
\begin{longtable}{|l|c|r|c|c|c|c|c|c|c|c|c|}
\caption{\small
\hsize=16.50cm Results of \AMORE\ with one of the parameters fixed at 
one sigma from its original value. 
These tests provide an
indication about the influence of one parameter on the retrieval
of the remaining parameters. 
In particular, how the remaining parameters balance this mis-match
by moving away from their optimal value. See
Sect.~\ref{desc_test4} and \ref{results_test4} for
additional details}
\label{fixwrong_tbl}\\
\hline
parameter & offset & model & $f_A$ & log d & A$_V$ & 
$\log t_{low}$ & $\log t_{hgh}$ & [Z]$_{low}$ & [Z]$_{hgh}$ &
$\alpha$ & $\beta$\\
\endfirsthead
\caption{continued}\\
\hline
parameter & offset & model & $f_A$ & log d & A$_V$ & 
$\log t_{low}$ & $\log t_{hgh}$ & [Z]$_{low}$ & [Z]$_{hgh}$ &
$\alpha$ & $\beta$\\
\hline
\endhead
\hline
\endfoot
\hline
log d    & $-1\sigma$ & 9  &  0.21045	 &  3.90307  &  0.008  &  9.70661  &  9.95123  & \minus0.54348  &  0.59806  & 2.318  & 4.573\\
	 &	& 14 &  0.29182	 &  3.90307  &  0.005  &  9.92758  &  9.95781  & \minus0.64641  &  0.10745  & 2.368  & 0.062\\
	 &	& 22 &  0.30663	 &  3.90307  &  0.055  &  9.90867  &  9.93700  & \minus0.62003  &  0.13004  & 2.332  & 0.373\\
	 &	& 34 &  0.27406	 &  3.90307  &  0.009  &  9.93190  &  9.95224  & \minus0.64172  &  0.09083  & 2.376  & 0.329\\
	 &	& 40 &  0.30508	 &  3.90307  &  0.031  &  9.88057  &  9.94765  & \minus0.56412  &  0.23885  & 2.330  & 1.482\\
	 &	& 52 &  0.35736	 &  3.90307  &  0.051  &  9.89002  &  9.94201  & \minus0.61187  &  0.18455  & 2.331  & 1.072\\
\hline
         & $+1\sigma$   & 9  &  0.29273	 &  3.90960  &  0.028  &  9.90776  &  9.93160  & \minus0.64116  &  0.10185  & 2.333  & 0.378\\
	 &	& 14 &  0.33350	 &  3.90960  &  0.007  &  9.89160  &  9.94261  & \minus0.62176  &  0.20013  & 2.336  & 1.201\\
	 &	& 22 &  0.29174	 &  3.90960  &  0.017  &  9.90814  &  9.94062  & \minus0.63653  &  0.11803  & 2.337  & 0.279\\
	 &	& 34 &  0.15593	 &  3.90960  &  0.000  &  9.82433  & 10.13280  & \minus0.74638  &  0.00763  & 2.346  & \minus1.894\\
	 &	& 40 &  0.33700	 &  3.90960  &  0.008  &  9.89589  &  9.94669  & \minus0.60483  &  0.15343  & 2.335  & 0.771\\
	 &	& 52 &  0.26929	 &  3.90960  &  0.032  &  9.91164  &  9.93459  & \minus0.64348  &  0.07215  & 2.340  & 0.145\\
\hline
A$_V$	 & $+1\sigma$	& 9  & - & - & - & - & - & - & - & - & -\\
	 &	& 14 &  0.27049	 &  3.89442  &  0.014  &  9.79866  &  9.96608  & \minus0.52682  &  0.49361  & 2.358  & 4.064\\
	 &	& 22 &  0.28352	 &  3.89378  &  0.014  &  9.87464  &  9.97332  & \minus0.55944  &  0.30339  & 2.359  & 1.855\\
	 &	& 34 &  0.28608	 &  3.89345  &  0.014  &  9.88671  &  9.97642  & \minus0.54707  &  0.30071  & 2.376  & 1.973\\
	 &	& 40 &  0.26382	 &  3.89346  &  0.014  &  9.81948  &  9.96660  & \minus0.52576  &  0.46184  & 2.358  & 3.610\\
	 &	& 52 &  0.31545	 &  3.89363  &  0.014  &  9.90003  &  9.97313  & \minus0.56649  &  0.23400  & 2.372  & 1.349\\
\hline
log t$_{low}$ & $-1\sigma$ & 9  & 0.27236  & 3.90040  & 0.040  & 9.85386  & 9.94757 & \minus0.55836  & 0.30935  & 2.323  & 1.896\\
              & & 14 & 0.27545  & 3.89466  & 0.019  & 9.85386  & 9.97024  & \minus0.52819  & 0.37310  & 2.368  & 2.888\\
              & & 22 & 0.27529  & 3.89648  & 0.018  & 9.85386  & 9.96558 & \minus0.54638  & 0.35495  & 2.350  & 2.384\\
              & & 34 & 0.16883  & 3.89375  & 0.029  & 9.85386  & 10.08302 & \minus0.47457  & 0.01306  & 2.400  & \minus0.280\\
              & & 40 & 0.28900  & 3.89953  & 0.033  & 9.85386  & 9.95405  & \minus0.56448  & 0.30320  & 2.324  & 1.779\\
              & & 52 & 0.26942  & 3.89372  & 0.047  & 9.85386  & 9.96388  & \minus0.55833  & 0.34280  & 2.356  & 2.939\\
\hline
              & $+1\sigma$ & 9 & 0.26839  & 3.89058  & 0.041  & 9.95232  & 9.95626 & \minus0.54291  & 0.11856  & 2.399  & 0.517\\
              & & 14   &  0.27352  & 3.88923  & 0.037  & 9.95232  & 9.96263 & \minus0.68969  & 0.08041  & 2.401 & \minus0.347\\
              & & 22   &  0.27555  & 3.88867  & 0.037  & 9.95232  & 9.95915 & \minus0.59030  & 0.06926  & 2.399 & \minus0.175\\
              & & 34   &  0.29222  & 3.88687  & 0.054  & 9.95232  & 9.95543 & \minus0.60299  & 0.07026  & 2.398 & \minus0.115\\
              & & 40   &  0.26972  & 3.88983  & 0.045  & 9.95232  & 9.95314 & \minus0.61176  & 0.08503  & 2.400  & 0.007\\
              & & 52   &  0.29660  & 3.89064  & 0.046  & 9.95232  & 9.95592 & \minus0.65754  & 0.08096  & 2.401  & 0.003\\
\hline
log t$_{high}$ & $-1\sigma$ & 9 & 0.21567  & 3.90212  & 0.093  & 9.90420  & 9.90749 & \minus0.59379  & 0.10497  & 2.306  & 0.206\\
              & & 14   &  0.25127  & 3.90319  & 0.111  & 9.90723  &
              9.90749 & \minus0.68463  & 0.09099  & 2.318  & 0.272\\
              & & 22   &  0.23271  & 3.90019  & 0.114  & 9.90634  &
              9.90749 & \minus0.69157  & 0.08733  & 2.316  & 0.167\\
              & & 34   &  0.24395  & 3.90526  & 0.113  & 9.89882  &
              9.90749 & \minus0.64179  & 0.09370  & 2.307  & 0.470\\
              & & 40   &  0.23796  & 3.90273  & 0.106  & 9.90270  &
              9.90750 & \minus0.63701  & 0.10850  & 2.306  & 0.378\\
              & & 52   &  0.19091  & 3.90474  & 0.105  & 9.85263  &
              9.90750 & \minus0.54484  & 0.26544  & 2.274  & 2.304\\
\hline
              & $+1\sigma$ & 9 & 0.18966  & 3.88133 & \minus0.001  & 9.84694 &
              10.00986 & \minus0.51375  & 0.40765  & 2.432  & 3.306\\
              & & 14   &  0.22621  & 3.88298  & \minus0.001 & 9.91537 &
              10.00986 & \minus0.50616  & 0.26567  & 2.437  & 2.059\\
              & & 22   &  0.21599  & 3.88275  & 0.000  & 9.73977 &
              10.00986 & \minus0.51424  & 0.51570  & 2.413  & 4.449\\
              & & 34   &  0.20975  & 3.88419  & \minus0.001 & 9.91194 &
              10.00985 & \minus0.51870  & 0.26155  & 2.427  & 2.046\\
              & & 40   &  0.19191  & 3.88460  & 0.013  & 9.79535 &
              10.00985 & \minus0.52199  & 0.43616  & 2.405  & 3.653\\
              & & 52   &  0.21132  & 3.88557  & 0.002  & 9.90671 &
              10.00987 & \minus0.51864  & 0.27274  & 2.422  & 2.044\\
\hline
log Z$_{low}$ & $-1\sigma$ & 9 & 0.32236  & 3.89424  & 0.060  & 9.92490  &
              9.94949 & \minus0.65235  & 0.09935  & 2.364  & 0.023\\
              &  & 14  & 0.19541  & 3.89205  & 0.085  & 9.80620 &
              9.95304  & \minus0.65235  & 0.44787  & 2.367  & 4.219\\
              &  &  22 & 0.19646  & 3.89803  & 0.068  & 9.80556 &
              9.95382 & \minus0.65235  & 0.48659  & 2.368  & 4.525\\
              &  &  34 & 0.19203  & 3.89730  & 0.066  & 9.76582 &
              9.94810 & \minus0.65235  & 0.48742  & 2.316  & 4.023\\
              &  &  40  & 0.32867  & 3.89484 &  0.038 &  9.90919 &
              9.95602 & \minus0.65235  & 0.18180  & 2.356  & 0.601\\
              &  &  52 & 0.28939  & 3.89843  & 0.035  & 9.88752 &
              9.95511 & \minus0.65235  & 0.22274  & 2.337  & 0.836\\
\hline
              & $+1\sigma$ & 9 & 0.30963  & 3.89302  & 0.019  & 9.89547  &
              9.97009 & \minus0.55177  & 0.26207  & 2.372  & 1.833\\
              & & 14 &  0.32883 &  3.89295 &  0.022 &  9.91000 &
              9.97315 & \minus0.55177 &  0.19130 &  2.365 &  0.778\\
              & & 22 &  0.32293 &  3.89110 &  0.024 &  9.91831 &
              9.96435 & \minus0.55177 &  0.18106 &  2.369 &  0.671\\
              & & 34 &  0.26376 &  3.89625 &  0.024 &  9.86434 &
              9.96369 & \minus0.55177 &  0.37255 &  2.361 &  2.928\\
              & & 40 &  0.33593 &  3.89840 &  0.032 &  9.90057 &
              9.96348 & \minus0.55177 &  0.17853 &  2.349 &  0.885\\
              & & 52 &  0.32988 &  3.89341 &  0.033 &  9.90827 &
              9.97162 & \minus0.55177 &  0.20534 &  2.376 &  1.416\\
\hline
log Z$_{high}$ & $-1\sigma$ & 9 & 0.29612  & 3.89604  & 0.077  & 9.92807  &
              9.93582 & \minus0.66338  & 0.04239  & 2.356 & \minus0.232\\
              & & 14 &  0.29094 &  3.89708 &  0.082 &  9.92897 &
              9.93400 & \minus0.67421 &  0.04239 &  2.359 & \minus0.139\\
              & & 22 &  0.27692 &  3.89364 &  0.066 &  9.94038 &
              9.95069 & \minus0.66032 &  0.04239 &  2.387 & \minus0.337\\
              & & 34 &  0.28980 &  3.89736 &  0.083 &  9.92808 &
              9.93087 & \minus0.63329 &  0.04239 &  2.357 & \minus0.095\\
              & & 40 &  0.16160 &  3.90196 &  0.043 &  9.82928 &
              10.10576 & \minus0.79270 &  0.04238 &  2.333 & \minus1.676\\
              & & 52 &  0.30348 &  3.89643 &  0.078 &  9.92740 &
              9.93508 & \minus0.66677 &  0.04239 &  2.355 & \minus0.238\\
\hline
              & $+1\sigma$ & 9 & 0.26702  & 3.88982  & 0.031  & 9.88500  &
              9.97615 & \minus0.56217  & 0.30979  & 2.384  & 2.280\\
              & & 14 &  0.29357 &  3.89392 &  0.012 &  9.88192 &
              9.97494 & \minus0.53776 &  0.30979 &  2.371 &  2.129\\
              & & 22 &  0.30635 &  3.89366 &  0.032 &  9.87176 &
              9.96383 & \minus0.53473 &  0.30979 &  2.351 &  2.194\\
              & & 34 &  0.25823 &  3.89389 &  0.046 &  9.87585 &
              9.96245 & \minus0.53400 &  0.30979 &  2.357 &  2.578\\
              & & 40 &  0.28417 &  3.89886 &  0.028 &  9.85863 &
              9.95559 & \minus0.57966 &  0.30979 &  2.333 &  1.926\\
              & & 52 &  0.28883 &  3.89501 &  0.020 &  9.88098 &
              9.96677 & \minus0.55136 &  0.30979 &  2.367 &  2.284\\
\hline	
$\alpha$ & $-1\sigma$	& 9  &  0.24524	 &  3.90004  &  0.053  &  9.85726  &  9.94126  & \minus0.56638  &  0.28025  & 2.316  & 1.932\\
	 & 	& 14 &  0.28147	 &  3.90272  &  0.020  &  9.87188  &  9.94707  & \minus0.54698  &  0.24944  & 2.316  & 1.230\\
	 & 	& 22 &  0.27797	 &  3.89896  &  0.042  &  9.86368  &  9.94580  & \minus0.54982  &  0.26102  & 2.316  & 1.511\\
	 & 	& 34 &  0.22427	 &  3.89644  &  0.032  &  9.82350  &  9.94534  & \minus0.50532  &  0.45532  & 2.316  & 3.679\\
	 & 	& 40 &  0.27857	 &  3.90056  &  0.042  &  9.86375  &  9.94608  & \minus0.59271  &  0.26364  & 2.316  & 1.294\\
	 & 	& 52 &  0.28640	 &  3.90139  &  0.039  &  9.85871  &  9.94768  & \minus0.58962  &  0.27721  & 2.316  & 1.438\\
\hline
         & $+1\sigma$  & 9   &  0.23946  &  3.89295  &  0.062  &  9.87606  &  9.96018  & \minus0.57093  &  0.35127  & 2.384  & 3.499\\
	 &	& 14 &  0.25586	 &  3.89172  &  0.014  &  9.86873  &  9.97973  & \minus0.53518  &  0.35320  & 2.384  & 2.919\\
	 &	& 22 &  0.28357	 &  3.89109  &  0.021  &  9.91357  &  9.97110  & \minus0.56888  &  0.24524  & 2.384  & 1.488\\
	 &	& 34 &  0.29934	 &  3.89146  &  0.023  &  9.91497  &  9.97387  & \minus0.58765  &  0.23086  & 2.384  & 1.461\\
	 &	& 40 &  0.25505	 &  3.89153  &  0.009  &  9.80529  &  9.97968  & \minus0.52454  &  0.51555  & 2.384  & 4.820\\
	 &	& 52 &  0.30437	 &  3.89128  &  0.026  &  9.90426  &  9.97836  & \minus0.55141  &  0.23875  & 2.384  & 1.620\\
\hline							
$\beta$	 & $-1\sigma$ & 9  &  0.29110	 &  3.89717  &  0.083  &  9.93098  &  9.93961  & \minus0.68243  &  0.02188  & 2.365	& \minus0.397\\
	 &	& 14 &  0.17266	 &  3.89314  &  0.001  &  9.83767  &  10.08181 & \minus0.41912  &  0.06629  & 2.372  & \minus0.397\\
	 &	& 22 &  0.30211	 &  3.89619  &  0.084  &  9.91729  &  9.94002  & \minus0.68522  &  0.06730  & 2.348  & \minus0.397\\
	 &	& 34 &  0.16366	 &  3.89491  &  0.015  &  9.81889  &  10.08661 & \minus0.50001  &  0.09684  & 2.363  & \minus0.397\\
	 &	& 40 &  0.16472	 &  3.89974  &  0.018  &  9.80958  &  10.07264 & \minus0.54158  &  0.11992  & 2.357  & \minus0.397\\
	 &	& 52 &  0.29684	 &  3.89592  &  0.074  &  9.91808  &  9.93138  & \minus0.66681  &  0.08493  & 2.339  & \minus0.397\\
\hline
         & $+1\sigma$   & 9  &  0.26953  &  3.89638  &  0.045  &  9.86719  &  9.95699  & \minus0.57054  &  0.30738  & 2.346  & 2.397\\
	 &	& 14 &  0.29111	 &  3.89398  &  0.023  &  9.87948  &  9.97371  & \minus0.54141  &  0.30286  & 2.371  & 2.397\\
	 &	& 22 &  0.28842	 &  3.89548  &  0.029  &  9.86193  &  9.95692  & \minus0.53941  &  0.33240  & 2.341  & 2.397\\
	 &	& 34 &  0.27336	 &  3.89607  &  0.048  &  9.88719  &  9.95851  & \minus0.56455  &  0.27131  & 2.355  & 2.397\\
	 &	& 40 &  0.26195	 &  3.89766  &  0.038  &  9.87053  &  9.94793  & \minus0.58042  &  0.30908  & 2.345  & 2.397\\
	 &	& 52 &  0.27328	 &  3.89034  &  0.022  &  9.87427  &  9.97430  & \minus0.54963  &  0.33427  & 2.375  & 2.397\\
\hline
\end{longtable}
}
\end{document}